\documentclass{llncs}
\usepackage[utf8]{inputenc} %
\usepackage{amsfonts,amssymb,amsmath}
\usepackage{enumitem}
\usepackage{listings,xspace,xcolor}
\usepackage{url}
\newcommand{\framac}{\textsc{Frama-C}\xspace}
\newcommand{\esapi}{\textsc{ESAPI}\xspace}

\newcommand{\acsl}{\textsc{acsl}\xspace}

\newcommand{\Wp}{\textsc{Wp}\xspace}

\newcommand{\Whythree}{\textsc{Why3}\xspace}
\newcommand{\Coq}{\textsc{Coq}\xspace}

\lstdefinelanguage{pretty-ACSL}{%
  escapechar={},
  breaklines=false,   
  literate=
   {==}{{$==$}}2
   {==>}{{$\Rightarrow$}}1
   {integer\ i}{{i$\,\in \mathbb{Z}\,$}}4
   {integer\ j}{{j$\,\in \mathbb{Z}\,$}}4
   {integer\ k}{{k$\,\in \mathbb{Z}\,$}}4
   {integer\ m}{{m$\,\in \mathbb{Z}\,$}}4
   {integer\ l}{{l$\,\in \mathbb{Z}\,$}}4
   {\\forall}{{$\forall$}}1
   {\\exists}{{$\exists$}}1
   {integer}{{$\mathbb{Z}$}}1
   {real}{{$\mathbb{R}$}}1
   {&&}{{$\wedge$}}1
   {||}{{$\vee$}}1
   {!=}{{$\neq$}}1
   {<}{{$<$}}1
   {<=}{{$\le~$}}1
   {>}{{$>$}}1
   {>=}{{$\ge~$}}1
   {<==>}{{$\Longleftrightarrow$}}1,
  morekeywords={assert,assigns,assumes,axiom,axiomatic,behavior,behaviors,
    boolean,breaks,complete,continues,data,decreases,disjoint,ensures,
    exit_behavior,ghost,global,inductive,invariant,lemma,logic,loop,
    model,predicate,relational,reads,requires,sizeof,strong,struct,terminates,
    type,union,variant,uchar,byte,typically,\\result,\\old,\\at,\\valid,
    \\separated,\\nothing,Pre,Post,Here,\\sum,\\numof,\\call,\\from},
  alsoletter={\\},
  morecomment=[l]{//}
}
\lstnewenvironment{pretty-codeACSL}{\lstset{language=pretty-ACSL,stepnumber=0}}{\smallskip}
\lstnewenvironment{pretty-codeACSL.smaller}{\lstset{language=pretty-ACSL,stepnumber=0,basicstyle=\scriptsize\ttfamily,}}{\smallskip}
\lstdefinelanguage{ACSL}{%
  escapechar={},
  literate=,
  breaklines=false,
  morekeywords={assert,assigns,assumes,axiom,axiomatic,behavior,behaviors,
    boolean,breaks,complete,continues,data,decreases,disjoint,ensures,
    exit_behavior,ghost,global,inductive,invariant,lemma,logic,loop,
    model,relational,predicate,reads,requires,sizeof,strong,struct,terminates,
    type,union,variant,uchar,byte,typically,\\result,\\old,\\at,\\valid,
    \\separated,\\nothing,Pre,\\exists,\\forall,\\sum,\\numof},
  alsoletter={\\},
  morecomment=[l]{//}
}
\lstMakeShortInline[language=ACSL]"
\lstnewenvironment{codeACSL}{\lstset{language=ACSL,stepnumber=0}}{\smallskip}
\lstnewenvironment{ACSL.small.figure}{\lstset{language=pretty-ACSL,basicstyle=\small\ttfamily,belowskip=0pt,aboveskip=0pt}}{}
\lstnewenvironment{ACSL.small}{\lstset{language=pretty-ACSL,basicstyle=\small\ttfamily,belowskip=6pt,aboveskip=6pt}}{}
\lstnewenvironment{ACSL.smaller.figure}{\lstset{language=pretty-ACSL,basicstyle=\scriptsize\ttfamily,belowskip=0pt,aboveskip=0pt}}{}
\lstnewenvironment{ACSL.smaller}{\lstset{language=pretty-ACSL,basicstyle=\scriptsize\ttfamily,belowskip=6pt,aboveskip=6pt}}{}
\lstdefinestyle{pretty-c}{language={[ANSI]C},%
  alsolanguage=pretty-ACSL,%
  moredelim={*[l]{//}},%
  deletecomment={[s]{/*}{*/}},
  moredelim={*[l]{//@}},%
}
\lstdefinestyle{c}{language={[ANSI]C},%
  alsolanguage=ACSL,%
  moredelim={*[l]{//}},%
  deletecomment={[s]{/*}{*/}},
  moredelim={*[l]{//@}},%
}
\title{Towards Formal Verification of\\a TPM Software Stack
}
\titlerunning{Towards Formal Verification of a TPM Software Stack}
\author{
  Yani Ziani \inst{1,2}${}^{[0009-0000-8540-1273]}$  \and
  Nikolai Kosmatov\inst{1}${}^{[0000-0003-1557-2813]}$  \and \\
  Fr\'{e}d\'{e}ric Loulergue\inst{2}${}^{[0000-0001-9301-7829]}$  \and \\
  Daniel Gracia P\'erez \inst{1}${}^{[0000-0002-5364-8244]}$  \and \\
  Téo Bernier \inst{1}${}^{[0009-0003-4834-7126]}$}
\institute{
  Thales Research \& Technology, Palaiseau, France \\
  \email{\{yani.ziani,nikolai.kosmatov,daniel.gracia-perez,teo.bernier\}@thalesgroup.com} \\
  \and
  {Université d’Orléans, INSA Centre Val de Loire, LIFO EA 4022}, France \\
  \email{frederic.loulergue@univ-orleans.fr}
}
\makeatletter%
\begin{document}
\maketitle
\begin{abstract}
The Trusted Platform Module (TPM) is a cryptoprocessor designed to protect
integrity and security of modern computers. Communications with the TPM
go through the TPM Software Stack (TSS), 
a popular implementation of which is the open-source library \emph{tpm2-tss}.
Vulnerabilities in its code could allow attackers to recover sensitive 
information and take control of the system.
This paper describes a case study on formal verification 
of tpm2-tss using the \framac verification platform.
Heavily based on linked lists and complex data structures,
the library code appears to be highly challenging for the verification tool.
We present several issues and limitations we faced, illustrate them with
examples and present solutions that allowed us to verify
functional properties and the absence of runtime errors for a representative 
subset of functions.
We describe verification results and desired tool improvements 
necessary to achieve a full formal verification of the target code.
 \end{abstract}
\section{Introduction}
\label{sec:intro}
\noindent
The \emph{Trusted Platform Module} (TPM)~\cite{TPM20}
has become a key security component in modern computers.
The TPM is a cryptoprocessor designed to protect
integrity of the architecture and ensure security of encryption keys
stored in it. The operating system and applications communicate with the TPM
through a set of APIs called \emph{TPM Software Stack} (TSS).
A popular implementation of the TSS is the open-source library \emph{tpm2-tss}.
It is highly critical: vulnerabilities in its code could allow attackers 
to recover sensitive information and take control of the system.
Hence, it is important to formally verify that 
the library respects its specification and does not contain 
runtime errors, often leading to security vulnerabilities, for instance, exploiting
buffer overflows or invalid pointer accesses. 
Formal verification of this library is the  main motivation of this work. 
This target is new and highly ambitious for deductive verification: the library code is 
very large for a formal verification project (over 120,000 lines of C code).
It is also highly 
complex, heavily based on 
complex data structures (with multiple levels of
nested structures and unions), low-level code,
calls to external (e.g. cryptography) libraries,
linked lists and dynamic memory allocation.

In this paper we present a first case study on formal verification of tpm2-tss
using the \framac verification platform~\cite{KKP2015:FAC}.
We focus on a subset of functions involved in storing an encryption key
in the TPM, one of the most critical features of the TSS. 
We verify both functional properties and the absence of runtime errors. 
The functions are annotated in the \acsl specification language~\cite{ACSL}.
Their verification with  \framac  currently
faces several limitations of the tool, such as its capacity to reason 
about complex data structures, dynamic memory allocation,
linked lists and their separation 
from other data.
We have managed to overcome these limitations after minor  
simplifications  and adaptations of the code.
In particular, 
we replace dynamic allocation with \lstinline|calloc| by another allocator 
(attributing preallocated memory cells) that we implement, specify and verify.
We adapt a recent work on verification 
of linked lists~\cite{BKL2019:SAC} to our case study,
add new lemmas and prove them in the \Coq proof assistant~\cite{coq}.
We identify some deficiencies in the 
new \framac--\Coq extraction for lists (modified since~\cite{BKL2019:SAC}),
adapt it for the proof and suggest improvements. 
We illustrate all issues and solutions on a simple illustrative 
example
while the (slightly adapted) real-life functions 
annotated in \acsl and fully proved in \framac
are available online as a companion artifact\footnote{Available (with the illustrative example, all necessary lemmas and their proofs) on \url{https://doi.org/10.5281/zenodo.8273295}.}. 
Finally, we identify desired extensions and improvements of the verification tool.

\vspace{-2mm}
\paragraph{Contributions.} The contributions of this paper include the following:
\begin{itemize}
\item specification and formal verification in \framac of 
a representative subset of functions of the tpm2-tss library
(slightly adapted for verification);
\item presentation of main issues we faced during their verification with
an illustrative example,
and description of solutions and workarounds we found;
\item proof in \Coq of all necessary lemmas (including some new ones) %
related to linked lists, realized for the new version of \framac--\Coq extraction;
\item a list of necessary enhancements of \framac to achieve a
complete formal verification of the tpm2-tss library.
\end{itemize}

\vspace{-2mm}
\paragraph{Outline.}
The paper is organized as follows. Section~\ref{sec:framac} presents \framac.
Section~\ref{sec:tss} 
introduces the  TPM, its software stack and the tpm2-tss library.
Sections~\ref{sec:dyn_memory} and~\ref{sec:memory_manage} present
issues and solutions related, resp., to memory allocation and memory management.
Necessary lemmas are discussed in Sect.~\ref{sec:lemmas}. 
Section~\ref{sec:results} describes our verification results. 
Finally, Sect.~\ref{sec:related_work} and~\ref{sec:discussion}  present 
related work and a conclusion with necessary tool improvements.

\section{\framac Verification Platform}
\label{sec:framac}

\framac~\cite{KKP2015:FAC} is an open-source verification platform for C code, which contains various plugins built around a kernel providing basic services for source-code 
analysis. It offers \acsl (ANSI/ISO C Specification Language)~\cite{ACSL}, a formal specification 
language for C, that allows users to specify functional properties of programs 
in the form of \emph{annotations}, such as assertions or function 
contracts. A function contract basically consists of pre- and postconditions (stated, resp., by  
\lstinline'requires' and \lstinline'ensures' clauses) expressing properties that must hold, resp.,
before and after a call to the function. 
It also includes an \lstinline{assigns} clause listing (non-local) 
variables and memory locations that \textit{can} 
be modified by the function.
While useful built-in predicates and logic functions are provided to handle 
properties such as pointer validity or memory separation for 
example, \acsl also supplies the user %
with 
different ways to define predicates and logic functions.

\framac offers \Wp, a plugin for deductive verification.
Given a C program annotated in \acsl, \Wp generates the corresponding 
proof obligations (also called verification conditions) that can be 
proved either by \Wp or, via the \Whythree 
platform~\cite{FilliatreP13}, by SMT solvers or an 
interactive proof assistant like \Coq~\cite{coq}.  
To ensure the absence of runtime errors (RTE), \Wp can automatically add
necessary assertions via a dedicated option, and try to prove them as well. 

Our choice to use \framac/\Wp 
is due to its capacity to perform deductive verification
of industrial C code with successful verification case studies~\cite{DHK2021:FM}
and the fact that it is currently the only tool for C source code verification 
recognized by ANSSI, the French Common Criteria certification body, as an acceptable 
formal verification technique for the highest  certification levels EAL6--EAL7~\cite{DjoudiHKKOMFF_ERTS22}.

\section{The TPM Software Stack and the tpm2-tss Library}
\label{sec:tss}

This section briefly presents the Trusted Platform Module (TPM), its software
stack and the implementation we chose to study: the tpm2-tss library.
Readers can refer to the TPM specification~\cite{TPM20} and
reference books as~\cite{AC2015} for more detail.

\paragraph{TPM Software Stack.}
The TPM is a standard conceived by the Trusted Computing Group
(TCG)\footnote{\url{https://trustedcomputinggroup.org/}} for a passive secure
cryptoprocessor designed to protect secure hardware from software-based
threats.
At its base, a TPM is implemented as a discrete cryptoprocessor chip, attached
to the main processor chip and designed to perform cryptographic
operations.
However, it can also be implemented as part of
the firmware of a regular processor or a software component.

Nowadays, the TPM is well known for its usage in regular PCs to ensure integrity
and to provide a secure storage for the keys used to encrypt the disk with
\textit{Bitlocker} and \textit{dm-crypt}.
However, it can be (and is actually) used  to provide other cryptographic
services to the Operating System (OS) or applications.
For that purpose, the TCG defines the TPM Software Stack (TSS), a set of
specifications to provide standard APIs to access the functionalities and
commands of the TPM, regardless of the hardware, OS, or
environment used.

The TSS APIs provide different levels of complexity, from the Feature API
(FAPI) for simple and common cryptographic services to the System API (SAPI)
for a one-to-one mapping to the TPM services and commands providing greater
flexibility but complexifying its usage.
In between lies the Enhanced System API (ESAPI) providing SAPI-like
functionalities but with slightly limited flexibility.
Other TSS APIs complete the previous ones for common operations like
data formatting and connection with the software or hardware TPM.

The TSS APIs, as any software component or the TPM itself, can have
vulnerabilities\footnote{Like CVE-2023-22745 and
CVE-2020-24455, documented on \url{www.cve.org}.} that attackers can exploit to recover sensitive data
communicated with the TPM or take control of the system.
We study the verification of one of the implementations of the
TSS, tpm2-tss, starting more precisely with its
implementation of the ESAPI.
\label{sec:tpm2-tss}

\paragraph{ESAPI Layer of tpm2-tss.}
The ESAPI layer %
provides functions for decryption and encryption, managing session data and policies,
thus playing an essential role in the TSS. 
It is very large (over 50,000 lines of C) 
and is mainly split into two parts: the API part containing
functions in a one-to-one correspondence with TPM commands (for instance, the \lstinline'Esys_Create'
function of the TSS will correspond to --- and call --- the \lstinline'TPM2_Create' command of the TPM),
and the back-end containing the core of that layer's functionalities.
Each API function will call several functions of the back-end to carry out various operations
on command parameters, before invoking the lower layers and finally the TPM.

The ESAPI layer relies on a notion of context (\lstinline'ESYS_CONTEXT')
containing all data the layer needs to store between calls,
so it does not need to maintain a global state.
Defined for external applications as an opaque structure,
the context includes, according to the documentation, data needed to communicate
to the TPM, metadata for each TPM resource, and state information.
The specification, however, does not impose any precise data structure:
it is up to the developer to provide a suitable definition.
Our target implementation uses complex data structures and linked lists.
\section{Dynamic Memory Allocation}
\label{sec:dyn_memory}
\paragraph{Example Overview.}
We illustrate our verification case study with a simplified version of some library
functions manipulating linked lists. The illustrative example is split into Fig.~\ref{fig:logic-lists-def}--\ref{fig:marshal_memcpy}
that will be explained below step-by-step.
Its full code being available in the companion artifact,
we omit in this paper some less significant definitions
and assertions which are not mandatory to understand the paper
(but we preserve line numbering of the full example for convenience of the reader).
This example is heavily simplified to fit the paper, yet it is representative for
most issues we faced (except the complexity of data structures).
It contains a main list manipulation function, \lstinline'getNode'
(\lstinline'esys_GetResourceObject' in the real code), used to search for a resource
in the list of resources and return it if it is found, or to create and add
it using function \lstinline'createNode'
(\lstinline'esys_CreateResourceObject' in the real code) if not.

Figure~\ref{fig:logic-lists-def} provides the linked list structure as well as
logic definitions used to handle logic lists in specifications.
Our custom allocator (used by \lstinline'createNode') is
defined  in Fig.~\ref{fig:calloc-def}.
Figure~\ref{fig:ctx_separations} defines a (simplified) context and additional logic definitions
to handle pointer separation and memory freshness.
The search function
is shown in Fig.~\ref{fig:getNode_1} 
and \ref{fig:getNode_rewrite}.
As it is often done, some \acsl notation (e.g. \verb'\forall', \verb'integer',  \verb'==>',  \verb'<=', \verb'!=')
is pretty-printed (resp., as \lstinline|\forall|, \lstinline|integer|, \lstinline|==>|, \lstinline|<=|, \lstinline|!=|).
In this section, we detail Fig.~\ref{fig:logic-lists-def}--\ref{fig:ctx_separations}.

\begin{figure}[tb]
    \begin{scriptsize}
        \dots
        \vspace{-2mm}
\begin{lstlisting}[firstnumber=11,firstline=11,lastline=15,language=pretty-ACSL,basicstyle=\scriptsize\ttfamily]
typedef struct NODE_T {
  uint32_t      handle;   // the handle used as reference
  RESOURCE      rsrc;     // the metadata for this rsrc
  struct NODE_T *   next; // next node in the list
} NODE_T; // linked list of resource
\end{lstlisting}
        \vspace{-3mm}
        \dots
        \vspace{-2mm}
\begin{lstlisting}[firstnumber=25,firstline=25,lastline=44,language=pretty-ACSL,basicstyle=\scriptsize\ttfamily]
/*@
  predicate ptr_sep_from_list{L}(NODE_T* e, \list<NODE_T*> ll) =
    \forall integer n; 0 <= n < \length(ll) ==> \separated(e, \nth(ll, n));
  predicate dptr_sep_from_list{L}(NODE_T** e, \list<NODE_T*> ll) =
    \forall integer n; 0 <= n < \length(ll) ==> \separated(e, \nth(ll, n));
  predicate in_list{L}(NODE_T* e, \list<NODE_T*> ll) =
    \exists integer n; 0 <= n < \length(ll) && \nth(ll, n) == e;
  predicate in_list_handle{L}(uint32_t out_handle, \list<NODE_T*> ll) =
    \exists integer n; 0 <= n < \length(ll) && \nth(ll, n)->handle == out_handle;
  inductive linked_ll{L}(NODE_T *bl, NODE_T *el, \list<NODE_T*> ll) {
    case linked_ll_nil{L}: \forall NODE_T *el; linked_ll{L}(el, el, \Nil);
    case linked_ll_cons{L}: \forall NODE_T *bl, *el, \list<NODE_T*> tail;
      (\separated(bl, el) && \valid(bl) && linked_ll{L}(bl->next, el, tail) &&
      ptr_sep_from_list(bl, tail)) ==>
        linked_ll{L}(bl, el, \Cons(bl, tail));
  }
  predicate unchanged_ll{L1, L2}(\list<NODE_T*> ll) =
    \forall integer n; 0 <= n < \length(ll) ==>
      \valid{L1}(\nth(ll,n)) && \valid{L2}(\nth(ll,n)) &&
      \at((\nth(ll,n))->next, L1) == \at((\nth(ll,n))->next, L2);
\end{lstlisting}        \vspace{-2mm}
        \dots
        \vspace{-2mm}
\begin{lstlisting}[firstnumber=48,firstline=48,lastline=60,language=pretty-ACSL,basicstyle=\scriptsize\ttfamily]
  axiomatic Node_To_ll {
    logic \list<NODE_T*> to_ll{L}(NODE_T* beg, NODE_T* end) 
      reads {node->next | NODE_T* node; \valid(node) &&
                           in_list(node, to_ll(beg, end))};
    axiom to_ll_nil{L}: \forall NODE_T *node; to_ll{L}(node, node) == \Nil;
    axiom to_ll_cons{L}: \forall NODE_T *beg, *end;
      (\separated(beg, end) && \valid{L}(beg) &&
      ptr_sep_from_list{L}(beg, to_ll{L}(beg->next, end))) ==>
        to_ll{L}(beg, end) == \Cons(beg, to_ll{L}(beg->next, end));
  }
*/

#include "lemmas_node_t.h"
\end{lstlisting}
        \vspace{-2mm}
    \end{scriptsize}
    \caption{Linked list and logic definitions.}
    \label{fig:logic-lists-def}
\end{figure}

\paragraph{Lists of Resources.}
Lines 11--15 of Fig.~\ref{fig:logic-lists-def} show a simplified definition
of the linked list of resources used in the ESAPI layer of the library. %
Each node of the list consists of a structure containing a handle used as a reference
for this node, a resource to be stored inside, and a pointer to the next element.
The handle is supposed to be unique\footnote{This uniqueness is currently not yet specified 
in the \acsl contracts.}.
In our example, a resource structure (omitted in Fig.~\ref{fig:logic-lists-def})
is assumed to contain only a few fields of relatively simple types.
The real code uses a more extensive and complex definition (with
several levels of nested structures and unions),
covering all possible types of TPM resources.
While it does add some complexity %
to prove certain
properties (as some of them may require to completely unfold
all resource substructures),
it does not introduce new pointers that may
affect memory separation properties, so our example remains representative
of the real code regarding linked lists and separation properties.

In particular, we need to ensure that the resource list is well-formed --- that is,
it is not circular, and does not contain any overlap between nodes --- and
stays that way throughout the layer.
To accomplish that, we use and adapt the logic definitions from~\cite{BKL2019:SAC},
given on lines 26--44, 48--57 of Fig.~\ref{fig:logic-lists-def}.
To prove the code, we need to manipulate linked lists and segments of linked lists.
Lines 48--57 define the \textit{translating function} \lstinline'to_ll' that
translates a C list defined by a \lstinline'NODE_T' pointer into the corresponding
\acsl logic list of (pointers to) its nodes.
By convention, the last element \lstinline|end| is not included into the resulting logic list.
It can be either \lstinline|NULL|
for a full linked list,
or a non-null pointer to a node for a \emph{linked list segment} which stops just before that node.
Lines 34--40 show the \textit{linking predicate} \lstinline'linked_ll'
establishing the equivalence between a C linked list and an \acsl logic list.
This inductive definition includes memory separation between nodes, validity of access for each node,
as well as the notion of reachability in linked lists.
In \acsl, given two pointers \lstinline|p| and \lstinline|q|,
\lstinline|\valid(p)| states that \lstinline|*p| can be safely read and written,
while \lstinline|\separated(p,q)| states that the referred memory locations \lstinline|*p| and \lstinline|*q|
do not overlap (i.e. all their bytes are disjoint).

Lines 26--29 provide predicates to handle separation between a list pointer (or double pointer) and a full list.
\lstinline|\nth(l,n)| and \lstinline|\length(l)| denote, resp.,
the \lstinline|n|-th element of logic list \lstinline|l|
and the length of \lstinline|l|.
The predicate \lstinline'unchanged_ll' in lines 41--44 states
that between two labels (i.e. program points) \lstinline'L1' and \lstinline'L2',
all list elements in a logic list refer to a valid memory location at both points,
and that their respective next fields retain the same value. It is used
to maintain the structure of the list throughout the code.
Line 60 includes lemmas %
necessary to conduct the proof, further discussed in Sec.~\ref{sec:lemmas}.

\paragraph{Lack of Support for Dynamic Memory Allocation.}
As mentioned above, per the TSS specifications, the ESAPI layer
does not maintain a global state between calls to TPM commands.
The library code uses contexts with linked lists of TPM resources, so list nodes
need to be dynamically allocated at runtime.
The \acsl language provides clauses to handle memory allocations: in particular,
\lstinline'\allocable{L}(p)' states that a pointer \lstinline'p'
refers to the base address of an unallocated memory block,
and \lstinline'\fresh{L1,L2}(p, n)' indicates that p refers
to the base address of an unallocated block at label \lstinline'L1', and to
an allocated memory block of size \lstinline'n' at label \lstinline'L2'.
Unfortunately, while the \framac/\Wp  memory model\footnote{that is, intuitively, the way in which program variables and memory locations are internally represented and manipulated by the tool.} is able to handle dynamic allocation
(used internally to manage local variables), these clauses are not currently supported.
{Without allocability and freshness, proving goals involving validity
or separation between a newly allocated node and any
other pointer is impossible.}

\paragraph{Static Memory Allocator.}
To circumvent that issue,
we define in Fig.~\ref{fig:calloc-def}
a bank-based static allocator \lstinline'calloc_NODE_T'
that replaces calls to \lstinline'calloc' used in the real-life code.
It attributes preallocated cells, following some existing implementations (like the memb module of Contiki~\cite{ManganoMembCRISIS2016}).
Line 63 defines a node bank, that is, a static array of nodes of size \lstinline'_alloc_max'.
Line 64 introduces an allocation index we use to track
the next allocable node and to determine whether an allocation is possible.
Predicate \lstinline'valid_rsrc_mem_bank' on line 66 states a validity condition for the bank:
\lstinline'_alloc_idx' must always be between
0 and \lstinline'_alloc_max'. It is equal to the upper bound if all nodes have been allocated.
Predicates lines 67--73 specify separation
between a logic list of nodes (resp., a pointer or a double pointer to a
node) and the allocable part of the heap, and is used later on to simulate memory freshness.

\begin{figure}[tb]
    \begin{scriptsize}
\begin{lstlisting}[firstnumber=62,firstline=62,lastline=74,language=pretty-ACSL,basicstyle=\scriptsize\ttfamily]
#define _alloc_max 100
static NODE_T _rsrc_bank[_alloc_max];  // bank used by the static allocator
static int _alloc_idx = 0;  // index of the next rsrc node to be allocated 
/*@ 
  predicate valid_rsrc_mem_bank{L} = 0 <= _alloc_idx <= _alloc_max;
  predicate list_sep_from_allocables{L}(\list<NODE_T*> ll) =
    \forall int i; _alloc_idx <= i < _alloc_max ==> 
                        ptr_sep_from_list{L}(&_rsrc_bank[i], ll);
  predicate ptr_sep_from_allocables{L}(NODE_T* node) =
    \forall int i; _alloc_idx <= i < _alloc_max ==> \separated(node, &_rsrc_bank[i]);
  predicate dptr_sep_from_allocables{L}(NODE_T** p_node) =
    \forall int i; _alloc_idx <= i < _alloc_max ==> \separated(p_node, &_rsrc_bank[i]);
*/
\end{lstlisting}
        \vspace{-2mm}
        \dots
        \vspace{-2mm}
\begin{lstlisting}[firstnumber=76,firstline=76,lastline=79,language=pretty-ACSL,basicstyle=\scriptsize\ttfamily]
/*@
  requires valid_rsrc_mem_bank;
  assigns _alloc_idx, _rsrc_bank[\old(_alloc_idx)];
  ensures valid_rsrc_mem_bank; 
\end{lstlisting}
        \vspace{-2mm}
        \dots
        \vspace{-2mm}
\begin{lstlisting}[firstnumber=89,firstline=89,lastline=111,language=pretty-ACSL,basicstyle=\scriptsize\ttfamily]
  behavior allocable:
    assumes 0 <= _alloc_idx < _alloc_max;
    
    ensures _alloc_idx == \old(_alloc_idx) + 1;
    ensures \result == &(_rsrc_bank[ _alloc_idx - 1]);
    ensures \valid(\result);
    ensures zero_rsrc_node_t( *(\result) );
    ensures \forall int i; 0 <= i < _alloc_max && i != \old(_alloc_idx) ==> 
            _rsrc_bank[i] == \old(_rsrc_bank[i]); 
  disjoint behaviors; complete behaviors;
*/
NODE_T *calloc_NODE_T()
{
  static const RESOURCE empty_RESOURCE;
  if(_alloc_idx < _alloc_max) {
    _rsrc_bank[_alloc_idx].handle = (uint32_t) 0;
    _rsrc_bank[_alloc_idx].rsrc = empty_RESOURCE;
    _rsrc_bank[_alloc_idx].next = NULL;
    _alloc_idx += 1;
    return &_rsrc_bank[_alloc_idx - 1];
  }
  return NULL;
}
\end{lstlisting}
    \end{scriptsize}
    \caption{Allocation bank and static allocator.}
    \label{fig:calloc-def}
\end{figure}

Lines 76--99 show a part of the function contract for the allocator defined on lines 100--111.
The validity of the bank should be true before and after the function execution
(lines 77, 79).
Line 78 specifies the variables the function is allowed to modify.
The contract is specified using several cases (called \emph{behaviors}).
Typically, a behavior considers a subset of possible input states (respecting its \lstinline'assumes' clause) and defines specific postconditions that must be respected for this subset of inputs. 
In our case, the provided behaviors are complete (i.e. cover all states allowed by the function precondition) 
and their corresponding subsets are disjoint  (line 98). 
We show only one behavior (lines 89--97)
describing a successful allocation (when an allocable node exists, as stated on line 90).
Postconditions on lines 92--93 ensure the tracking index is incremented by one,
and that the returned pointer points to  the first allocable block.
While this fact is sufficient to deduce the validity clause on line 94,
we keep the latter as well (and it is actually expected for any allocator).
In the same way, lines 96--97 specify that the nodes of the bank
other than the newly allocated block have not been modified\footnote{This property 
is partly redundant with the assigns clause on line 78 but its presence 
facilitates the verification.}.

Currently, \framac/\Wp does not offer a memory model able to handle byte-level assignments in C objects.
To represent as closely as possible the fact that allocated memory is initialized
to zero by a call to \lstinline'calloc' in the real-life code,
we initialize each field of the allocated node to zero
(see the C code on lines 104--106 and the postcondition on line 95).

\paragraph{Contexts, Separation Predicates and Freshness.}
In the target library (and in our illustrative example),
pointers to nodes are not passed directly as function
arguments, but stored in a context variable, and a pointer to the context
is passed as a function argument.
Lines 113--116 of Fig.~\ref{fig:ctx_separations} define a simplified context structure, comprised of an \lstinline'int' and a \lstinline'NODE_T' pointer to the head of a linked list of resources.

Additional predicates to handle memory separation
and memory freshness are defined on lines 118--132.
In particular, the \lstinline'ctx_sep_from_list' predicate on lines 118--119
specifies memory separation between a \lstinline'CONTEXT' pointer
and a logic list of nodes.
Lines 120--121 define separation between such a pointer and allocables nodes in the bank.

\begin{figure}[tb]
    \begin{scriptsize}
\begin{lstlisting}[firstnumber=113,firstline=113,lastline=134,language=pretty-ACSL,basicstyle=\scriptsize\ttfamily]
typedef struct CONTEXT {
  int placeholder_int;
  NODE_T *rsrc_list;    
} CONTEXT;
/*@
  predicate ctx_sep_from_list(CONTEXT *ctx, \list<NODE_T*> ll) =  
    \forall integer i; 0 <= i < \length(ll) ==> \separated(\nth(ll, i), ctx);
  predicate ctx_sep_from_allocables(CONTEXT *ctx) =
    \forall int i; _alloc_idx <= i < _alloc_max ==> \separated(ctx, &_rsrc_bank[i]);
    
  predicate freshness(CONTEXT * ctx, NODE_T ** node) = 
    ctx_sep_from_allocables(ctx)
    && list_sep_from_allocables(to_ll(ctx->rsrc_list, NULL))
    && ptr_sep_from_allocables(ctx->rsrc_list)
    && ptr_sep_from_allocables(*node)
    && dptr_sep_from_allocables(node);
    
  predicate sep_from_list{L}(CONTEXT * ctx, NODE_T ** node) = 
    ctx_sep_from_list(ctx, to_ll{L}(ctx->rsrc_list, NULL))
    && dptr_sep_from_list(node, to_ll{L}(ctx->rsrc_list, NULL));
*/
\end{lstlisting}
    \end{scriptsize}
    \caption{Context and predicates to handle separation from a list and memory freshness.}
    \label{fig:ctx_separations}
\end{figure}

In C, a successful dynamic allocation of a memory block implies its \emph{freshness},
that is, the separation between the newly allocated block (typically located on the heap)
and all pre-existing memory locations (on the heap, stack or
static storages).
As this notion of freshness is currently not supported by \framac/\Wp,
we have to simulate it in another way.
Our allocator returns a cell in a static array,
so other global variables --- as well as local
variables declared within the scope of a function ---
will be separated from the node bank.
To obtain a complete freshness within the scope of a function,
we need to maintain separation between the allocable part of the
bank and other memory locations accessible through pointers.
In our illustrative example, pointers come from arguments including a pointer
to a \lstinline'CONTEXT' object
(and pointers accessible from it)
and a double pointer to a \lstinline'NODE_T' node.
This allows us to define a predicate to handle freshness
in both function contracts.

The \lstinline'freshness' predicate on lines 123--128 of
Fig.~\ref{fig:ctx_separations}  specifies memory separation
between known pointers within the scope of our functions and the allocable part of the
bank, using separation predicates previously defined on
lines 120--121, and on lines 67--73 of Fig \ref{fig:calloc-def}.
This predicate will become unnecessary as soon as
dynamic allocation is fully supported by \framac/\Wp.

In the meanwhile, a static allocator with an additional separation
predicate simulating freshness
provides a reasonable solution to verify the target library.
Since no specific constraint is assumed in our contracts
on the position of previously allocated list nodes already added to the list,
the verification uses a specific position in the bank
only for the newly allocated node.
The fact that the newly allocated node does not
become valid during the allocation (technically, being part of the bank,
it was valid in the sense of \acsl already before) is compensated in our contracts
by the freshness predicate stating that the new node
--- as one the allocable nodes --- was not used in the list before the allocation
(cf. line 310 in Fig.~\ref{fig:getNode_1}).
We expect that the migration
from our specific allocator
to a real-life dynamic allocator --- with
a more general contract ---
will be very easy to perform, as soon as necessary features are supported by \framac.

Similarly, the \lstinline'sep_from_list' predicate on lines
130--132 specifies separation between the context's linked list and
known pointers, using predicates on lines
118--119, and on lines 28--29 of Fig \ref{fig:logic-lists-def}.

\section{Memory Management}
\label{sec:memory_manage}
This section presents how we use the definitions introduced 
in Sec. \ref{sec:dyn_memory} to prove selected \esapi functions 
involving linked lists. 
We also identify 
separation issues related to limitations of the Typed memory model of \Wp,
as well as a way to manage memory to 
overcome such issues.
In this section, we detail Fig.~\ref{fig:getNode_1}--\ref{fig:marshal_memcpy}.

\paragraph{The Search Function.}
Figure~\ref{fig:getNode_1} provides the search operation \lstinline'getNode' 
with  a partial contract 
illustrating  functional and memory safety properties we aim to 
verify and judge necessary for the proof at a larger scale. 
Some proof-guiding annotations (assertions, loop contracts)
have been skipped %
for readability, 
but the code is preserved (mostly with the same line numbers).
The arguments include a context, a handle to search and a double pointer 
for the returned node. 

Lines 380--416 perform the search of a node by its handle: variable 
\lstinline'temp_node' iterates over the nodes of the resource list, and the node is 
returned if its handle is equal to the searched one (in which case, the 
function returns 616 for success).
    
Lines 420--430 convert the resource handle to a TPM one, call the creation 
function to allocate a new node and add it to the list as its new head with the given 
handle if the allocation was successful (and return \lstinline'833' if not). 
The new node is returned by  \lstinline'createNode' in \lstinline'temp_node_2' 
(again via a double pointer).
    
Lines 435--462 perform some modifications on the content of the newly allocated node, 
without affecting the structure of the list. An error code is returned
in case of a failure, and 1611 (with the allocated node in \lstinline'*node') 
otherwise.     
Lines 450--451, 453--454 and 461 provide some assertions to propagate 
information to the last return clause of the function, attained 
in case of the successful addition of the new element to the  list.

Compared to the real-life code, we have introduced 
anonymous blocks on lines 380--416 and 422--452 (which are not 
semantically necessary and were not present in the original code), 
as well as two local variables 
\lstinline'tmp_node' and \lstinline'tmp_node2' instead of only one.
We explain these code adaptations below.

\begin{figure}[t!]
    \begin{scriptsize}
\begin{lstlisting}[firstnumber=309,firstline=309,lastline=310,language=pretty-ACSL,basicstyle=\scriptsize\ttfamily]
/*@
  requires valid_rsrc_mem_bank{Pre} && freshness(ctx, node);
\end{lstlisting} 
        \vspace{-4.2mm}
\begin{lstlisting}[firstnumber=313,firstline=313,lastline=313,language=pretty-ACSL,basicstyle=\scriptsize\ttfamily]
  requires linked_ll(ctx->rsrc_list, NULL, to_ll(ctx->rsrc_list, NULL));
\end{lstlisting} 
        \vspace{-4.2mm} 
\begin{lstlisting}[firstnumber=317,firstline=317,lastline=317,language=pretty-ACSL,basicstyle=\scriptsize\ttfamily]
  requires sep_from_list(ctx, node) && \separated(node, ctx);
\end{lstlisting} 
        \vspace{-3mm}
        \dots
        \vspace{-2mm}
\begin{lstlisting}[firstnumber=321,firstline=321,lastline=321,language=pretty-ACSL,basicstyle=\scriptsize\ttfamily]
  ensures valid_rsrc_mem_bank && freshness(ctx, node);
\end{lstlisting} 
        \vspace{-3mm}
        \dots
        \vspace{-2mm}
\begin{lstlisting}[firstnumber=325,firstline=325,lastline=326,language=pretty-ACSL,basicstyle=\scriptsize\ttfamily]
  behavior handle_in_list:
    assumes in_list_handle(rsrc_handle, to_ll(ctx->rsrc_list, NULL));
\end{lstlisting} 
        \vspace{-3mm}
        \dots
        \vspace{-2mm}
\begin{lstlisting}[firstnumber=332,firstline=332,lastline=333,language=pretty-ACSL,basicstyle=\scriptsize\ttfamily]
    ensures unchanged_ll{Pre, Post}(to_ll(ctx->rsrc_list, NULL));
    ensures \result == 616;
\end{lstlisting} 
        \vspace{-3mm}
        \dots
        \vspace{-2mm}
\begin{lstlisting}[firstnumber=355,firstline=355,lastline=359,language=pretty-ACSL,basicstyle=\scriptsize\ttfamily]
  behavior handle_not_in_list_and_node_allocated:
    assumes !(in_list_handle(rsrc_handle, to_ll(ctx->rsrc_list, NULL)));
    assumes rsrc_handle <= 31U || (rsrc_handle \in {0x10AU, 0x10BU})
            || (0x120U <= rsrc_handle <= 0x12FU);
    assumes 0 <= _alloc_idx < _alloc_max;
\end{lstlisting}
        \vspace{-3mm}
        \dots
        \vspace{-2mm}
\begin{lstlisting}[firstnumber=369,firstline=369,lastline=375,language=pretty-ACSL,basicstyle=\scriptsize\ttfamily]
    ensures \old(ctx->rsrc_list) != NULL ==> 
            \nth(to_ll(ctx->rsrc_list, NULL), 1) == \old(ctx->rsrc_list);
    ensures linked_ll(ctx->rsrc_list, NULL, to_ll(ctx->rsrc_list, NULL)); 
    ensures sep_from_list(ctx, node);
    ensures \result == 1611;
  disjoint behaviors; complete behaviors;
*/
\end{lstlisting}        
        \vspace{-4.3mm}
\begin{lstlisting}[firstnumber=376,firstline=247,lastline=252,language=pretty-ACSL,basicstyle=\scriptsize\ttfamily]
int getNode(PSEUDO_CONTEXT *ctx, uint32_t rsrc_handle, NODE_T ** node) {
  /*@ assert linked_ll(ctx->rsrc_list, NULL, to_ll(ctx->rsrc_list, NULL));*/
  int r;
  uint32_t tpm_handle;
  { /* Block added to circumvent issues with the WP memory model */ 
    NODE_T *tmp_node;
\end{lstlisting}
        \vspace{-4.3mm}
\begin{lstlisting}[firstnumber=401,firstline=253,lastline=254,language=pretty-ACSL,basicstyle=\scriptsize\ttfamily]
    for (tmp_node = ctx->rsrc_list; tmp_node != NULL; 
        tmp_node = tmp_node->next) {
\end{lstlisting}
        \vspace{-4.3mm}
\begin{lstlisting}[firstnumber=405,firstline=255,lastline=255,language=pretty-ACSL,basicstyle=\scriptsize\ttfamily]
      if (tmp_node->handle == rsrc_handle){*node = tmp_node; return 616;}
\end{lstlisting}
        \vspace{-4.3mm}
\begin{lstlisting}[firstnumber=407,firstline=256,lastline=256,language=pretty-ACSL,basicstyle=\scriptsize\ttfamily]
    
\end{lstlisting}
        \vspace{-4.3mm}
\begin{lstlisting}[firstnumber=410,firstline=257,lastline=258,language=pretty-ACSL,basicstyle=\scriptsize\ttfamily]
    
    
\end{lstlisting}
        \vspace{-4.3mm}
\begin{lstlisting}[firstnumber=415,firstline=259,lastline=260,language=pretty-ACSL,basicstyle=\scriptsize\ttfamily]
    }
  }
\end{lstlisting}
        \vspace{-4.3mm}
\begin{lstlisting}[firstnumber=420,firstline=261,lastline=261,language=pretty-ACSL,basicstyle=\scriptsize\ttfamily]
  r = iesys_handle_to_tpm_handle(rsrc_handle, &tpm_handle);
\end{lstlisting}
        \vspace{-4.3mm}
\begin{lstlisting}[firstnumber=422,firstline=262,lastline=263,language=pretty-ACSL,basicstyle=\scriptsize\ttfamily]
  { /* Block added to circumvent issues with the WP memory model */
    NODE_T *tmp_node_2 = NULL;
\end{lstlisting}
        \vspace{-4.3mm}
\begin{lstlisting}[firstnumber=428,firstline=264,lastline=266,language=pretty-ACSL,basicstyle=\scriptsize\ttfamily]
    r = createNode(ctx, rsrc_handle, &tmp_node_2);
    /*@ assert sep_from_list(ctx, node);*/
    if (r == 833) {return r;};
\end{lstlisting}
        \vspace{-4.3mm}
\begin{lstlisting}[firstnumber=435,firstline=267,lastline=269,language=pretty-ACSL,basicstyle=\scriptsize\ttfamily]
    tmp_node_2->rsrc.handle = tpm_handle;
    tmp_node_2->rsrc.rsrcType = 0;
    size_t offset = 0;
\end{lstlisting}
        \vspace{-4.3mm}
\begin{lstlisting}[firstnumber=440,firstline=270,lastline=271,language=pretty-ACSL,basicstyle=\scriptsize\ttfamily]
    r = uint32_Marshal(tpm_handle, &tmp_node_2->rsrc.name.name[0],
                          sizeof(tmp_node_2->rsrc.name.name),&offset);
\end{lstlisting}
        \vspace{-4.3mm}
\begin{lstlisting}[firstnumber=443,firstline=272,lastline=273,language=pretty-ACSL,basicstyle=\scriptsize\ttfamily]
    if (r !=0) {return r;};
    tmp_node_2->rsrc.name.size = offset;
\end{lstlisting}
        \vspace{-4.3mm}
\begin{lstlisting}[firstnumber=449,firstline=274,lastline=279,language=pretty-ACSL,basicstyle=\scriptsize\ttfamily]
    *node = tmp_node_2;
    /*@ assert unchanged_ll{Pre, Here}(
               to_ll{Pre}(\at(ctx->rsrc_list, Pre), NULL));*/
  }   
  /*@ assert unchanged_ll{Pre, Here}(
             to_ll{Pre}(\at(ctx->rsrc_list, Pre), NULL));*/
\end{lstlisting}
        \vspace{-4.3mm}
\begin{lstlisting}[firstnumber=461,firstline=280,lastline=282,language=pretty-ACSL,basicstyle=\scriptsize\ttfamily]
  /*@ assert sep_from_list(ctx, node);*/
  return 1611;
}
\end{lstlisting}
    \end{scriptsize}
    \caption{The (slightly rewritten) search function, where some annotations are removed.}
    \label{fig:getNode_1}
\end{figure}

\paragraph{Contract of the Search Function.}
Lines 309--375 of Fig.~\ref{fig:getNode_1} provide a partial function contract, 
illustrating two behaviors of \lstinline'getNode': if the element was 
found by its handle in the list (cf. lines 325--326),
and if the element was not found at first, 
but was then successfully allocated and added (cf. lines 355--359), for each 
of them specific postconditions are stated.
For instance, for the latter behavior, 
lines 369--370 ensure that if a new node was successfully allocated and 
added to the list, the old head becomes the second element of the list,
while line 372 ensures the separation of known pointers from the new list.
We specify that the complete list of provided behaviors must be
complete and disjoint (line 374).

As global preconditions, we notably require  for the list to be well-formed 
(through the use of the linking predicate, cf. line 313), 
and  the validity of our bank 
and freshness of allocable nodes with respect to function arguments and 
global variables (cf. line 310). 
Line 317 requires 
memory separation of known pointers from the list of resources using the 
\lstinline'sep_from_list' predicate, and separation among known pointers using the  
\lstinline'\separated' predicate.

As a global postcondition, we require  that our bank stays valid, and that 
freshness of the (remaining) allocable nodes 
relatively to function arguments and global variables is 
maintained (cf. line 321). 
However, 
properties regarding the list itself --- such as the preservation of the list when it is 
not modified (line 332), or ensuring it remains well-formed after being modified (line 371) --- have to be 
issued to \acsl behaviors to be proved, due to the way 
how local variables are 
handled in the memory model of \Wp. 
The logic list properties are 
much more difficult for solvers to manipulate in global behaviors.

\paragraph{Memory Model Limitation: an Uprovable Property.}
Consider the assertion on line 377 of Fig.~\ref{fig:getNode_1}. 
Despite the presence of the same property as a precondition of the function
(line 313), currently this assertion cannot be proved by \Wp at the entry point
for the real-life version of the function. Basically, the real-life version can be obtained\footnote{another difference ---
removing variable \lstinline'tmp_node2' declared on line 423 
and using \lstinline'tmp_node' instead --- can be ignored in this context.} from 
Fig.~\ref{fig:getNode_1} by removing 
the curly braces on lines 380, 416, 422, 452.
This issue is due to a limitation of the \Wp memory model.

Indeed, for such an 
assertion (as in general for any annotation to be proved), \Wp generates a proof 
obligation, to be proved by either \Wp itself or by 
external provers via the \Whythree 
platform~\cite{FilliatreP13}. 
Such an obligation includes a representation of the current 
state of the program memory. In particular,  
pointers such as the resource list \lstinline'ctx->rsrc_list' (and by 
extension, any reachable node of the list) will be considered part of the heap. 
To handle the existence of a variable in memory --- should it be  the heap, 
the stack or the 
static segments --- \Wp uses an allocation table to express when memory 
blocks are used or 
freed, which is where the issue lies. For instance, on line 428 of 
Fig.~\ref{fig:getNode_1}, 
the \lstinline'temp_node_2' pointer has its address taken, and is 
considered as used locally due to \lstinline'requires' involving it in our function 
contract for \lstinline'createNode'. It is consequently transferred to the memory model, 
where it has to be allocated. 

Currently, the memory model of \Wp does not provide separated allocation tables for the 
heap, stack and static segments. Using \lstinline'temp_node_2' the way it is used on line 
428 changes the modification status of the allocation table, 
which is then considered as modified as a whole. This affects the status of 
other ''allocated'' (relatively to the memory model) variables as well, including 
(but not limited to) any reachable node of the list.

Therefore, the call to \lstinline'createNode' line 428 of Fig.~\ref{fig:getNode_1} 
in the real-life code that uses the address of a local pointer as a third argument
is sufficient to affect the status of the resource list on the scale of the entire 
function. As a result, the assertion  on line 377  is not proved.

\paragraph{A Workaround.}
As a workaround (found thanks to an
indication of the \Wp team) to the aforementioned issue, 
we use additional blocks and variable declarations. 
Figure~\ref{fig:getNode_rewrite} presents those minor rewrites (with line numbers
in alphabetical style to avoid confusion with the illustrative example). 
The left side 
illustrates the structure of the original C code, where the address of \lstinline'temp_node' is taken and used in the \lstinline'createNode' call 
on line j, and the same pointer is used to iterate on the list. 
On the right, we add additional blocks and a new pointer \lstinline'temp_node_2', 
initialized to  \lstinline'NULL' to match the previous iteration over the list. 
Each block defines a new  scope, outside of which the pointer 
used by \lstinline'createNode' will not exist and 
side-effect-prone allocations will not happen. 
It solves the issue.

\begin{figure}[tb]
\renewcommand*\thelstnumber{\alph{lstnumber}}	
    \noindent\begin{minipage}{.48\textwidth}
        \begin{lstlisting}[language=pretty-ACSL,basicstyle=\scriptsize\ttfamily]
int getNode(..., NODE_T ** node){ 	
  // list properties unprovable
  int r;
  
  NODE_T *tmp_node;
  ... // iterate over the list
    
    
    
  r = createNode(..., &tmp_node);
  ...
  *node = tmp_node;
    
  return 1611;
}
        \end{lstlisting}
    \end{minipage}\hfill
    \begin{minipage}{.48\textwidth}
        \begin{lstlisting}[language=pretty-ACSL,basicstyle=\scriptsize\ttfamily]
int getNode(..., NODE_T ** node){ 
  // list properties proved
  int r;
  { 
    NODE_T *tmp_node;
    ... // iterate over the list
  }  
  {
    NODE_T *tmp_node_2 = NULL;
    r = createNode(..., &tmp_node_2);
    ...
    *node = tmp_node_2;
  }
  return 1611;
}
        \end{lstlisting}
    \end{minipage}
    \caption{Comparison of the real-life code of \lstinline|getNode| (on the left) and its rewriting with additional blocks (on the right) for proving list properties.}
    \label{fig:getNode_rewrite}
\end{figure}

\paragraph{Additional Proof-Guiding Annotations.}
Additional annotations (mostly omitted in Fig.~\ref{fig:getNode_1})
include, as usual, loop contracts and a few assertions. 
Assertions can help the tool to establish necessary intermediate properties or
activate the application of relevant lemmas.
For instance, assertions of 
lines 450--451 and 453--454 help propagate information over the 
structure of the linked list (by its logic list representation) outside 
of each block, and finally to postconditions. 
Assertions on lines 429 and 461 help propagate separations 
from the list through the function and its anonymous blocks.
Some other intermediate assertions are needed to prove the unchanged nature of the list. 
Such additional assertions can be tricky to find in some cases and need
some experience.

\begin{figure}[tb]
    \begin{scriptsize}
\begin{lstlisting}[firstnumber=271,firstline=271,lastline=272,language=pretty-ACSL,basicstyle=\scriptsize\ttfamily]
/*@ 
  requires \valid(src) && \valid(dest + (0 .. sizeof(*src)-1));
\end{lstlisting}  
        \vspace{-3mm}
        \dots
        \vspace{-2mm}
\begin{lstlisting}[firstnumber=279,firstline=279,lastline=285,language=pretty-ACSL,basicstyle=\scriptsize\ttfamily]
*/
void memcpy_custom(uint8_t *dest, uint32_t *src) { 
  dest[3] = (uint8_t)(*src & 0xFF);
  dest[2] = (uint8_t)((*src >> 8) & 0xFF);
  dest[1] = (uint8_t)((*src >> 16) & 0xFF);
  dest[0] = (uint8_t)((*src >> 24) & 0xFF);
}
\end{lstlisting}  
        \vspace{-3mm}
        \dots
        \vspace{-2mm}
\begin{lstlisting}[firstnumber=298,firstline=298,lastline=299,language=pretty-ACSL,basicstyle=\scriptsize\ttfamily]
int uint32_Marshal(uint32_t in,uint8_t buff[],size_t buff_size,size_t *offset){ 
  size_t  local_offset = 0; 
\end{lstlisting}  
        \vspace{-3mm}
        \dots
        \vspace{-2mm}
\begin{lstlisting}[firstnumber=302,firstline=302,lastline=303,language=pretty-ACSL,basicstyle=\scriptsize\ttfamily]
  // memcpy(&buff[local_offset], &in, sizeof (in));
  memcpy_custom(&buff[local_offset], &in);  
\end{lstlisting}\vspace{-3mm}
        \dots
        \vspace{-2mm}
\begin{lstlisting}[firstnumber=306,firstline=306,lastline=306,language=pretty-ACSL,basicstyle=\scriptsize\ttfamily]
}
\end{lstlisting}
    \end{scriptsize}
    \caption{Definition for \lstinline|memcpy| replacement in marshal.}
    \label{fig:marshal_memcpy}
\end{figure}

\paragraph{Handling Pointer Casts.}
Another memory manipulation issue we have encountered  
comes from the function call on line 440 in \lstinline'getNode': after 
having been added to the resource list, the newly allocated node must have its 
name (or more precisely, the name of its resource) set from its TPM handle \lstinline'tpm_handle' (derived from the handle of the node by the function 
call on line 420). This is done through marshaling using 
the \lstinline'uint32_Marshal' 
function, 
partially shown on lines 298--306 of Fig.~\ref{fig:marshal_memcpy}, whose role is to 
store a 4-byte unsigned int (in this case, our TPM handle) in a flexible array of bytes 
(the name of the resource). 
The function 
calls \lstinline'memcpy' on (commented) line 302, which is the source of our issue
(a correct endianness being ensured by a previous byte swap in \lstinline'in').

For most functions of the standard libraries, \framac provides basic \acsl contracts to 
handle their use. However, for memory manipulation functions like \lstinline'memcpy', 
such contracts rely on pointer casts, whose support in \Wp is currently limited.
To circumvent this issue, we define our own memory copy function on 
lines 280--285: instead of directly copying the 4-byte unsigned int pointed by 
\lstinline'src' byte per byte using pointer casts 
using \lstinline'memcpy', we extract one-byte chunks 
using byte shifts and bitmasks (cf. lines 281--284, 303) without casts. 
Line 272 requires that both source and destination locations  
are valid, also without casts. 
This version is fully handled by \Wp.
Current contracts are sufficient for the currently considered
functional properties and the absence of runtime errors
(and we expect they will be easy to extend for more precise properties if needed).

\section{Lemmas}
\label{sec:lemmas}

When SMT solvers become inefficient (e.g. for inductive definitions),
it can be necessary to add lemmas to facilitate the proof.
These lemmas can then be directly instantiated by solvers, but 
proving them often requires to reason by induction, with an interactive proof assistant. 

The previous work using logic lists~\cite{BKL2019:SAC} defined and proved 
several lemmas using the \Coq proof assistant. 
We have added two new useful lemmas (defined in Fig.~\ref{fig:lemmas-new})
and used twelve of the previous ones to verify
both the illustrative example and the subset of real-life functions.
However, because the formalization of 
the memory models and various aspects of \acsl  changed between the 
version of \framac used in the previous work and the one we use, 
we could not reuse the 
proofs of these lemmas. While older \framac versions directly generated \Coq 
specifications, more recent \framac versions let \Whythree generate them. 
Even if the new translation 
is close to the previous one, the way logic lists are handled was modified 
significantly.

\begin{figure}[t!]
    \begin{scriptsize}
\begin{lstlisting}[numbers=none, firstline=33,lastline=40,language=pretty-ACSL,basicstyle=\scriptsize\ttfamily]
  lemma in_next_not_bound_in{L}: \forall NODE_T *bl, *el, *item, \list<NODE_T*> ll;
    linked_ll(bl, el, ll) ==> in_list(item, ll) ==> item->next != el ==>
      in_list(item->next, ll);
  lemma linked_ll_split_variant{L}: 
    \forall NODE_T *bl, *bound, *el, \list<NODE_T*> l1, l2;
    linked_ll(bl, el, l1 ^ l2) ==> l2 != \Nil ==>
    bound == \nth(l1 ^ l2, \length(l1 ^ l2) - \length(l2)) ==>
      linked_ll(bl, bound, l1) && linked_ll(bound, el, l2);
\end{lstlisting}
    \end{scriptsize}
    \caption{New lemmas proved in our verification work (in addition to those in~\cite{BKL2019:SAC}).}
    \label{fig:lemmas-new}
\end{figure}

In the past, \framac logic lists were translated into the lists \Coq offers in its 
standard library: an inductively defined type as usually found in functional programming 
languages such as OCaml and Haskell. Such types come with an induction principle that 
allows to reason by induction. Without reasoning inductively, it also offers the 
possibility to reason by case on lists: a list is defined either as empty, or as built 
with the \lstinline'cons' constructor.
In recent versions of \framac, \acsl logic lists are axiomatized as follows: two 
functions \lstinline'nil' and \lstinline'cons' are declared, as well as a few other 
functions on lists, including the length of a list (\lstinline'length'), the 
concatenation of two lists (\lstinline'concat'), and getting an element from a list given 
its position (\lstinline'nth').
However, there is no induction principle to reason by induction on lists, and because 
\lstinline'nil' and \lstinline'cons' are not constructors, it is not possible to reason 
by case on lists in this formalization. It is possible to test if a list is empty, but if 
not, we do not know that it is built with \lstinline'cons'.
Writing new recursive functions on such lists is also very difficult. Indeed, we only 
have \lstinline'nth' to observe a list, while the usual way to program functions on lists 
uses the head 
and the tail of a list for writing the 
recursive case.

Interestingly, when the hypotheses of our lemmas include a fact expressed using \lstinline'linked_ll', it is still possible to reason by case,
because this inductive predicate 
is translated into \Coq as an inductive predicate. Consequently, there are only two 
possible cases for the logic list: either it is empty, or it is built with 
\lstinline'cons'.
When such a hypothesis is missing, we axiomatized a \lstinline'tail' function, and a 
decomposition principle stating that a list is either \lstinline'nil' or 
\lstinline'cons'. These axioms are quite classic and can be implemented using a list type 
defined by induction.
We did not need an inductive principle on logic lists as either the lemmas did not require a 
proof by induction, or we reasoned inductively on 
the inductive predicate  \lstinline'linked_ll'. However, we 
proved such an induction principle using only the axioms we added.
It is thus available to prove some other 
lemmas provided in~\cite{BKL2019:SAC} --- not needed yet in our current
work --- that were proved by induction on lists.

Because of these changes, to prove all lemmas we need, 
we had to adapt all previous proof scripts, and in a few 
cases significantly. The largest proof scripts are about 100 lines long excluding our 
axioms, and the shortest takes a dozen lines. 
We suggest that the next versions of \framac 
come back to a concrete representation of lists. 
Thanks to our approach, we expect that
the required changes in our proofs of lemmas
will remain minimal: we will only have to prove the axioms introduced on 
\lstinline'tail' and our decomposition principle.

\section{Verification Results}
\label{sec:results}

Proof results, presented in Fig.~\ref{table_proof_results_smoke}, were obtained by running \framac 26.1 (Iron) on a desktop computer running Ubuntu 20.04.4 LTS, with an Intel(R) Core(TM) i5-6600 CPU @ 3.30 GHz, featuring 4 cores and 4 threads, with 16GB RAM. We ran \framac with options \lstinline'-wp-par 3' and \lstinline'-wp-timeout 30'. We used the Alt-Ergo v2.4.3 and CVC4 v1.8 solvers, via \Whythree v1.5.1.
Both functional properties and the absence of runtime errors (RTE) were proved.
Assertions to ensure the absence of runtime errors are automatically generated
by the \textsc{Rte} plugin of \framac (using the \lstinline'-wp-rte' option).
Functional properties include usual properties such as 
the fact 
that the well-formedness of the list is preserved, 
that a new resource  has been successfully added to the resource list, 
that the searched element is correctly found if present, etc. 

\begin{figure}[tb]
    \centering
    \scriptsize
    \begin{tabular}{|l | l | c | c || c|c|}
        \cline{3-6}
        \multicolumn{2}{c|}{} & \multicolumn{1}{c|}{User-provided} & \multicolumn{1}{c||}{RTE} & \multicolumn{2}{c|}{Total}          \\
        \multicolumn{2}{c|}{} & \multicolumn{1}{c|}{\acsl}         & \multicolumn{1}{c||}{}    & \multicolumn{2}{c|}{ }              \\
        \hline
        Code subset           & Prover         & $\#_{\mathrm{Goals}}$     & $\#_{\mathrm{Goals}}$      & $\#_{\mathrm{Goals}}$ & Time         \\
        \hline
        Illustrative       & Qed              & 105            & 18            & 123 (43.62\%)    &    \\
        example            & Script           & 1              & 0             & 1 (0.35\%)       &    \\
                           & SMT              & 137            & 21            & 158 (56.03\%)    &    \\
        \cline{2-6}
                           & All              & 243 (86.17\%)  & 39 (13.83\%)  & {\bf 282}        & {\bf 5m13s}  \\
        \hline
        Library            & Qed              & 274            & 38            & 312 (47.34\%)    &    \\
        code subset        & Script           & 5              & 0             & 5 (0.76\%)       &    \\
                           & SMT              & 311            & 31	           & 342 (51.90\%)    &    \\
        \cline{2-6}
                           & All              & 590 (89.53\%)  & 69 (10.47\%)  & {\bf 659}        & {\bf 18m07s} \\
        \hline
    \end{tabular}
    \caption{Proof results for the illustrative example and the real-life code.}
    \label{table_proof_results_smoke}
\end{figure}

In our illustrative example, 282 goals were proved in a total time of 5min13s with 56\% proved by SMT solvers, and the rest by the internal simplifier engine Qed of \Wp and one \Wp script. The maximum time to prove a goal was 20s.

Solutions to memory manipulation problems presented in this paper
were used on a larger verification study over 10 different functions of the
target library (excluding macro functions, and interfaces without code whose behaviors
needed to be modeled in \acsl), related to linked-list manipulations and some internal
ESAPI feasibility checks and operations (cryptographic operations excluded). Over 659 
goals proved in a total of 18m07s, 52\% were proved by SMT solvers
and 47\% by Qed. 
Only 5 \Wp proof scripts were used, when automatic proof either failed
or was too slow.
This shows a high level of automation achieved in our project,
in particular, thanks to carefully chosen predicates and lemmas
(which are usually tricky to find for the first time 
and can be useful in other similar projects).
The maximum time to prove a goal was
1min50s.

We also used smoke-tests to detect unexpected dead code or possible 
inconsistencies in the specification, and manually checked that no unexpected cases 
of those were detected.

As for the 14 lemmas we used, 11 are proved by \Coq using our scripts, and the remaining 3 directly by Alt-Ergo. Their proof takes 6 seconds in our configuration, with the maximum time to prove a goal being 650ms.

\section{Related Work}
\label{sec:related_work}
\paragraph{TPM related safety and security.} 
Various case studies centered around TPM uses have emerged over the last decade, often focusing on use cases relying on functionalities of the TPM itself. A recent formal analysis of the key exchange primitive of TPM 2.0~\cite{ZhangZ20} provides a security model to capture TPM protections on keys and protocols. Authors of~\cite{WangQYZF16} propose a security model for the cryptographic support commands in TPM 2.0, proved using the CryptoVerif tool. A model of TPM commands was used to formalize the session-based HMAC authorization and encryption mechanisms~\cite{ShaoQF18}. Such works focus on the TPM itself, but to the best of our knowledge, none of the previously published works aim at verifying the tpm2-tss library or any implementation of the TSS.

\paragraph{Linked lists and recursive data structures.} 
We use logical definitions from~\cite{BKL2019:SAC} to formalize and manipulate C linked lists as \acsl logic lists in our effort, while another approach~\cite{BKL2018:NFM} relies on a parallel view of a linked list via a companion ghost array. Both approaches were tested on the linked list module of the Contiki OS~\cite{DGV2004:LCN}, which relies on static allocations and simple structures. 
In this work we used a logic list based approach rather than a ghost code based approach
following the conclusions in~\cite{BKL2019:SAC}. 
Realized in SPARK, a deductive verification tool for a subset of the Ada language and also the name of this subset, the approach to the verification of black-red trees~\cite{DBLP:conf/nfm/DrossM17} is related to the verification of linked lists in \framac using ghost arrays including the auto-verification aspects~\cite{BLK2019:NFM}. However, the trees themselves were implemented using arrays as pointers have only been recently introduced in SPARK~\cite{DBLP:conf/cav/DrossK20}. Programs with pointers in SPARK are based on an ownership policy enforcing non-aliasing which makes their verification closer to Rust programs than C programs.

\paragraph{Formal verification for real-life code.}
Deductive verification on real-life code has been spreading in the last decades, with various verification case studies where bugs were often found by annotating and verifying the code~\cite{Hahnle2019}. Such studies include~\cite{Dordowsky2015}, providing feedback on the authors' experience of using \acsl and \framac on a real-world example. Authors of~\cite{DHK2021:FM} managed a large scale formal verification of global security properties on the C code of the JavaCard Virtual Machine. SPARK was used in the verification of a TCP Stack~\cite{DBLP:conf/secdev/CluzelGMZ21}. Authors of~\cite{DBLP:conf/fm/LeinenbachS09} highlight some issues specific to the verification of the Hyper-V hypervisor, and how they can be solved with VCC, a deductive verification tool for C.

\section{Conclusion and Future Work}
\label{sec:discussion}
This paper presents a first case study on formal verification of the tpm2-tss library,
a popular implementation of the TPM Software Stack.
Making the bridge between the TPM and applications,
this library is highly critical: to take advantage of security guarantees
of the TPM, its deductive verification is highly desired.
The library code is very complex and challenging for verification tools.

We have presented our verification results for a subset of 10 functions of the
ESAPI layer of the library that we verified with \framac.
We have described current limitations of the verification tool
and temporary solutions we used to address them.
We have proved all necessary lemmas (extending those of a previous case study for linked
lists~\cite{BKL2019:SAC}) in \Coq
using the most recent version of the \framac--\Coq translation
and identified some necessary improvements in handling logic lists.
Finally, we identified desired tool improvements to achieve a full
formal verification of the library: support of dynamic allocations and basic \acsl clauses to handle them, a memory model that works at byte level, and clearer separation of modification
statuses of variables between the heap, the stack, and static segments.
The real-life code was slightly simplified for verification,
but the logical behavior was preserved in the verified version. 
While the current real-life code cannot be verified without adaptations, 
we expect that it will become
provable 
as soon as those improvements  of the tool are implemented\footnote{Detailed
discussions of limitations and ongoing extensions of \framac can be found at \url{https://git.frama-c.com/pub/frama-c/}.}.

This work opens the way towards a full verification
of the tpm2-tss library.
Future work includes the verification of a larger subset of functions,
including lower-level layers and operations.
Specification and verification of specific security properties is
another future work direction.
Maintaining proofs for changing versions of tools and axiomatizations is also
an interesting research direction.
Finally, combining formally verified modules with modules
which undergo a partial verification (e.g. limited to the absence of runtime errors,
or runtime assertion checking of expected specifications on large test suites)
can be another promising work direction to increase
confidence in the security of the library.

\paragraph{Acknowledgment.}
Part of this work was supported by ANR (grants ANR-22-CE39-0014, %
ANR-22-CE25-0018%
) and French Ministry of Defense via a PhD grant of Yani Ziani.
We thank Allan Blanchard, Laurent Corbin and Lo\"{i}c Correnson
for useful discussions, and 
the anonymous referees for helpful comments.
\bibliographystyle{splncs03}
\bibliography{bibliography}
\newpage
\appendix
\section{Appendix: Supplementary Material}

This appendix 
presents the complete illustrative example.

\subsection{Complete Illustrative Example}
Figures~\ref{fig:self-contained-list-code-1}, \ref{fig:self-contained-list-code-2}, \ref{fig:self-contained-list-code-3}, \ref{fig:self-contained-list-code-4}, \ref{fig:self-contained-list-code-5}, \ref{fig:self-contained-list-code-6}, \ref{fig:self-contained-list-code-7}, \ref{fig:self-contained-list-code-8} give the complete
version of the illustrative example
(presented in Fig.~\ref{fig:logic-lists-def}--\ref{fig:marshal_memcpy} in the paper),
annotated in \acsl.
It was proved with \framac 26.1, \Whythree 1.5.1, Alt-Ergo 2.4.3 and CVC4 1.8. The command
used to run the proof is given at the end of the file.

Figure \ref{fig:self-contained-list-lemmas-1} provides the definition of the lemmas required to perform the proof.
The same lemmas are used for the illustrative example and the proved subset of the real-life code.
All necessary lemmas were proved with \Coq 8.16.1 (but other recent versions should also work).
The \Coq proof scripts and the instructions how to run the proof are available in the companion artifact.

\begin{figure}[ht!]
    \vspace{-4mm}
    \begin{scriptsize}
\begin{lstlisting}[firstnumber=1,firstline=1,lastline=67,language=pretty-ACSL,basicstyle=\scriptsize\ttfamily]
/********************** lemmas_node_t.h **********************/
/*@
  lemma linked_ll_in_valid{L}: \forall NODE_T *bl, *el, \list<NODE_T*> ll;
    linked_ll(bl, el, ll) ==> \forall integer n ; 0 <= n < \length(ll) ==>
      \valid(\nth(ll, n));
  lemma ptr_sep_from_nil{L}: \forall NODE_T* l;
    ptr_sep_from_list(l, \Nil);
  lemma ptr_sep_from_cons{L}: \forall NODE_T *e, *hd, \list<NODE_T*> l;
    ptr_sep_from_list(e, \Cons(hd, l)) <==>
      (\separated(hd, e) && ptr_sep_from_list(e, l));
  lemma dptr_sep_from_nil{L}:
    \forall NODE_T** l ; dptr_sep_from_list(l, \Nil);
  lemma linked_ll_all_separated{L}: \forall NODE_T *bl, *el, \list<NODE_T*> ll;
    linked_ll(bl, el, ll) ==> all_sep_in_list(ll);
  lemma linked_ll_unchanged_ll{L1, L2}: \forall NODE_T *bl, *el, \list<NODE_T*> ll;
    linked_ll{L1}(bl, el, ll) ==>
      unchanged_ll{L1, L2}(ll) ==> linked_ll{L2}(bl, el, ll);
  lemma linked_ll_to_ll{L}: \forall NODE_T *bl, *el, \list<NODE_T*> ll;
    linked_ll(bl, el, ll) ==> ll == to_ll(bl, el);
  lemma to_ll_split{L}:  \forall NODE_T *bl, *el, *sep, \list<NODE_T*> ll;
    ll != \Nil ==> linked_ll(bl, el, ll) ==> ll == to_ll(bl, el) ==>
      in_list(sep, ll) ==> ll == (to_ll(bl, sep) ^ to_ll(sep, el));
  lemma in_list_in_sublist: \forall NODE_T* e, \list<NODE_T*> rl, ll, l;
    (rl ^ ll) == l ==> (in_list(e, l) <==> (in_list(e, rl) || in_list(e, ll)));
  lemma linked_ll_end{L}: \forall NODE_T *bl, *el, \list<NODE_T*> ll;
    ll != \Nil ==> linked_ll(bl, el, ll) ==>
      \nth(ll, \length(ll)-1)->next == el;
  lemma linked_ll_end_separated{L}: \forall NODE_T *bl, *el, \list<NODE_T*> ll;
    linked_ll(bl, el, ll) ==> ptr_sep_from_list(el, ll);
  lemma linked_ll_end_not_in{L}: \forall NODE_T *bl, *el, \list<NODE_T*> ll;
    linked_ll(bl, el, ll) ==> !in_list(el, ll);
//new lemmas wrt. previous work on linked lists [Blanchard et al., SAC'19] 
  lemma in_next_not_bound_in{L}: \forall NODE_T *bl, *el, *item, \list<NODE_T*> ll;
    linked_ll(bl, el, ll) ==> in_list(item, ll) ==> item->next != el ==>
      in_list(item->next, ll);
  lemma linked_ll_split_variant{L}: 
    \forall NODE_T *bl, *bound, *el, \list<NODE_T*> l1, l2;
    linked_ll(bl, el, l1 ^ l2) ==> l2 != \Nil ==>
    bound == \nth(l1 ^ l2, \length(l1 ^ l2) - \length(l2)) ==>
      linked_ll(bl, bound, l1) && linked_ll(bound, el, l2);
*/
\end{lstlisting}
    \end{scriptsize}
    \vspace{-4mm}
    \caption{Lemmas used to prove the illustrative example and the subset of real-life code.}
    \label{fig:self-contained-list-lemmas-1}
\end{figure}

\begin{figure}[p]
    \begin{scriptsize}
\begin{lstlisting}[firstnumber=1,firstline=1,lastline=60,language=pretty-ACSL,basicstyle=\scriptsize\ttfamily]
#include <stdint.h>       // for uint types definitions
#include <string.h>       // for size_t definition
#include <byteswap.h>     // used in marshal
#define HOST_TO_BE_32(value) __bswap_32 (value) // swap endianness
typedef struct TPM2B_NAME { uint16_t size; uint8_t name[68];} TPM2B_NAME;
typedef struct {
  uint32_t      handle;   // handle used by TPM
  TPM2B_NAME      name;   // TPM name of the object
  uint32_t      rsrcType; // selector for resource type
} RESOURCE;
typedef struct NODE_T {
  uint32_t      handle;   // the handle used as reference
  RESOURCE      rsrc;     // the metadata for this rsrc
  struct NODE_T *   next; // next node in the list
} NODE_T; // linked list of resource
/*@
  predicate zero_tpm2b_name(TPM2B_NAME tpm2b_name) =
    tpm2b_name.size == 0 && \forall int i; 0 <= i  < 68 ==> tpm2b_name.name[i] == 0;
  predicate zero_resource(RESOURCE rsrc) =
    rsrc.handle == 0 && zero_tpm2b_name(rsrc.name) && rsrc.rsrcType == 0;
  predicate zero_rsrc_node_t(NODE_T node) =
    node.handle == 0 && zero_resource(node.rsrc) && node.next == \null;
*/
/************** Logic lists and linked lists definitions *************/
/*@
  predicate ptr_sep_from_list{L}(NODE_T* e, \list<NODE_T*> ll) =
    \forall integer n; 0 <= n < \length(ll) ==> \separated(e, \nth(ll, n));
  predicate dptr_sep_from_list{L}(NODE_T** e, \list<NODE_T*> ll) =
    \forall integer n; 0 <= n < \length(ll) ==> \separated(e, \nth(ll, n));
  predicate in_list{L}(NODE_T* e, \list<NODE_T*> ll) =
    \exists integer n; 0 <= n < \length(ll) && \nth(ll, n) == e;
  predicate in_list_handle{L}(uint32_t out_handle, \list<NODE_T*> ll) =
    \exists integer n; 0 <= n < \length(ll) && \nth(ll, n)->handle == out_handle;
  inductive linked_ll{L}(NODE_T *bl, NODE_T *el, \list<NODE_T*> ll) {
    case linked_ll_nil{L}: \forall NODE_T *el; linked_ll{L}(el, el, \Nil);
    case linked_ll_cons{L}: \forall NODE_T *bl, *el, \list<NODE_T*> tail;
      (\separated(bl, el) && \valid(bl) && linked_ll{L}(bl->next, el, tail) &&
      ptr_sep_from_list(bl, tail)) ==>
        linked_ll{L}(bl, el, \Cons(bl, tail));
  }
  predicate unchanged_ll{L1, L2}(\list<NODE_T*> ll) =
    \forall integer n; 0 <= n < \length(ll) ==>
      \valid{L1}(\nth(ll,n)) && \valid{L2}(\nth(ll,n)) &&
      \at((\nth(ll,n))->next, L1) == \at((\nth(ll,n))->next, L2);
  predicate all_sep_in_list(\list<NODE_T*> ll) =
    \forall integer n1, n2; (0 <= n1 < \length(ll) && 0 <= n2 < \length(ll) && n1 != n2) ==>
        \separated(\nth(ll, n1), \nth(ll, n2));
  axiomatic Node_To_ll {
    logic \list<NODE_T*> to_ll{L}(NODE_T* beg, NODE_T* end) 
      reads {node->next | NODE_T* node; \valid(node) &&
                           in_list(node, to_ll(beg, end))};
    axiom to_ll_nil{L}: \forall NODE_T *node; to_ll{L}(node, node) == \Nil;
    axiom to_ll_cons{L}: \forall NODE_T *beg, *end;
      (\separated(beg, end) && \valid{L}(beg) &&
      ptr_sep_from_list{L}(beg, to_ll{L}(beg->next, end))) ==>
        to_ll{L}(beg, end) == \Cons(beg, to_ll{L}(beg->next, end));
  }
*/

#include "lemmas_node_t.h"
\end{lstlisting}
    \end{scriptsize}
    \vspace{-4mm}
    \caption{Illustrative provable example of the adjusted tpm2-tss list manipulation code, part 1/8.}
    \label{fig:self-contained-list-code-1}
\end{figure}

\begin{figure}[p]
    \begin{scriptsize}
\begin{lstlisting}[firstnumber=61,firstline=61,lastline=111,language=pretty-ACSL,basicstyle=\scriptsize\ttfamily]

#define _alloc_max 100
static NODE_T _rsrc_bank[_alloc_max];  // bank used by the static allocator
static int _alloc_idx = 0;  // index of the next rsrc node to be allocated 
/*@ 
  predicate valid_rsrc_mem_bank{L} = 0 <= _alloc_idx <= _alloc_max;
  predicate list_sep_from_allocables{L}(\list<NODE_T*> ll) =
    \forall int i; _alloc_idx <= i < _alloc_max ==> 
                        ptr_sep_from_list{L}(&_rsrc_bank[i], ll);
  predicate ptr_sep_from_allocables{L}(NODE_T* node) =
    \forall int i; _alloc_idx <= i < _alloc_max ==> \separated(node, &_rsrc_bank[i]);
  predicate dptr_sep_from_allocables{L}(NODE_T** p_node) =
    \forall int i; _alloc_idx <= i < _alloc_max ==> \separated(p_node, &_rsrc_bank[i]);
*/
/***************************************************************************/
/*@
  requires valid_rsrc_mem_bank;
  assigns _alloc_idx, _rsrc_bank[\old(_alloc_idx)];
  ensures valid_rsrc_mem_bank; 
  
  behavior not_allocable:
    assumes _alloc_idx == _alloc_max;
    
    ensures _alloc_idx == _alloc_max;
    ensures \result == NULL;
    ensures _rsrc_bank == \old(_rsrc_bank);
    ensures \forall int i; 0 <= i < _alloc_max ==> 
              _rsrc_bank[i] == \old(_rsrc_bank[i]); 
  behavior allocable:
    assumes 0 <= _alloc_idx < _alloc_max;
    
    ensures _alloc_idx == \old(_alloc_idx) + 1;
    ensures \result == &(_rsrc_bank[ _alloc_idx - 1]);
    ensures \valid(\result);
    ensures zero_rsrc_node_t( *(\result) );
    ensures \forall int i; 0 <= i < _alloc_max && i != \old(_alloc_idx) ==> 
            _rsrc_bank[i] == \old(_rsrc_bank[i]); 
  disjoint behaviors; complete behaviors;
*/
NODE_T *calloc_NODE_T()
{
  static const RESOURCE empty_RESOURCE;
  if(_alloc_idx < _alloc_max) {
    _rsrc_bank[_alloc_idx].handle = (uint32_t) 0;
    _rsrc_bank[_alloc_idx].rsrc = empty_RESOURCE;
    _rsrc_bank[_alloc_idx].next = NULL;
    _alloc_idx += 1;
    return &_rsrc_bank[_alloc_idx - 1];
  }
  return NULL;
}
\end{lstlisting}
    \end{scriptsize}
    \vspace{-4mm}
    \caption{Illustrative provable example of the adjusted tpm2-tss list manipulation code, part 2/8.}
    \label{fig:self-contained-list-code-2}
\end{figure}

\begin{figure}[p]
    \begin{scriptsize}
\begin{lstlisting}[firstnumber=112,firstline=112,lastline=176,language=pretty-ACSL,basicstyle=\scriptsize\ttfamily]

typedef struct CONTEXT {
  int placeholder_int;
  NODE_T *rsrc_list;    
} CONTEXT;
/*@
  predicate ctx_sep_from_list(CONTEXT *ctx, \list<NODE_T*> ll) =  
    \forall integer i; 0 <= i < \length(ll) ==> \separated(\nth(ll, i), ctx);
  predicate ctx_sep_from_allocables(CONTEXT *ctx) =
    \forall int i; _alloc_idx <= i < _alloc_max ==> \separated(ctx, &_rsrc_bank[i]);
    
  predicate freshness(CONTEXT * ctx, NODE_T ** node) = 
    ctx_sep_from_allocables(ctx)
    && list_sep_from_allocables(to_ll(ctx->rsrc_list, NULL))
    && ptr_sep_from_allocables(ctx->rsrc_list)
    && ptr_sep_from_allocables(*node)
    && dptr_sep_from_allocables(node);
    
  predicate sep_from_list{L}(CONTEXT * ctx, NODE_T ** node) = 
    ctx_sep_from_list(ctx, to_ll{L}(ctx->rsrc_list, NULL))
    && dptr_sep_from_list(node, to_ll{L}(ctx->rsrc_list, NULL));
*/

/*@
  requires valid_rsrc_mem_bank && freshness(ctx, out_node);
  requires \valid(ctx);
  requires ctx->rsrc_list != NULL ==> \valid(ctx->rsrc_list);
  requires linked_ll(ctx->rsrc_list, NULL, to_ll(ctx->rsrc_list, NULL));
  requires sep_from_list(ctx, out_node);
  requires ptr_sep_from_list(*out_node, to_ll(ctx->rsrc_list, NULL));
  requires !(in_list_handle(esys_handle, to_ll(ctx->rsrc_list, NULL)));
  requires \valid(out_node) && \separated(ctx, out_node);
  requires *out_node != NULL ==> \valid(*out_node) && (*out_node)->next == NULL;
  assigns _alloc_idx, _rsrc_bank[_alloc_idx], ctx->rsrc_list, *out_node;
  ensures valid_rsrc_mem_bank && freshness(ctx, out_node);  
  ensures sep_from_list(ctx, out_node);  
  ensures unchanged_ll{Pre, Post}(to_ll{Pre}(\old(ctx->rsrc_list), NULL));
  ensures \result \in {1610, 833};
  
  behavior not_allocable:
    assumes _alloc_idx == _alloc_max;
    
    ensures _alloc_idx == _alloc_max;
    ensures \valid(ctx);
    ensures !(in_list_handle(esys_handle, to_ll(ctx->rsrc_list, NULL)));
    ensures ctx->rsrc_list == \old(ctx->rsrc_list);
    ensures *out_node == \old(*out_node);
    ensures unchanged_ll{Pre, Post}(to_ll(ctx->rsrc_list, NULL));
    ensures \result == 833;
  behavior allocated:
    assumes 0 <= _alloc_idx < _alloc_max;
    
    ensures _alloc_idx == \old(_alloc_idx) + 1;
    ensures in_list_handle(esys_handle, to_ll(ctx->rsrc_list, NULL));
    ensures \valid(ctx->rsrc_list) && *out_node == ctx->rsrc_list;
    ensures ctx->rsrc_list == &_rsrc_bank[_alloc_idx - 1]; 
    ensures ctx->rsrc_list->handle == esys_handle;
    ensures ctx->rsrc_list->next == \old(ctx->rsrc_list);
    ensures linked_ll(ctx->rsrc_list, NULL, to_ll(ctx->rsrc_list, NULL));
    ensures unchanged_ll{Pre, Post}(to_ll{Pre}(\old(ctx->rsrc_list), NULL));
    ensures \old(ctx->rsrc_list) != NULL ==>
        \nth(to_ll(ctx->rsrc_list, NULL), 1) == \old(ctx->rsrc_list);
    ensures \result == 1610;    
  disjoint behaviors; complete behaviors;
*/
\end{lstlisting}
    \end{scriptsize}
    \vspace{-4mm}
    \caption{Illustrative provable example of the adjusted tpm2-tss list manipulation code, part 3/8.}
    \label{fig:self-contained-list-code-3}
\end{figure}

\begin{figure}[p]
    \begin{scriptsize}
\begin{lstlisting}[firstnumber=177,firstline=177,lastline=242,language=pretty-ACSL,basicstyle=\scriptsize\ttfamily]
int createNode(CONTEXT * ctx, uint32_t esys_handle, NODE_T ** out_node){
//@ ghost pre_calloc:;
  //@ghost int if_id = 0;
  /*@ assert \separated(out_node, &_rsrc_bank[_alloc_idx]);*/ 
  /*@ assert \separated(ctx->rsrc_list, &_rsrc_bank[_alloc_idx]); */
  // NODE_T *new_head = calloc(1, sizeof(NODE_T));  /*library version*/
  /*@ assert list_sep_from_allocables(to_ll(ctx->rsrc_list, NULL)); */
  /*@ assert ptr_sep_from_list(&_rsrc_bank[_alloc_idx], to_ll{pre_calloc}(ctx->rsrc_list, NULL)); */
  /*@ assert ptr_sep_from_list(&_rsrc_bank[_alloc_idx], to_ll(ctx->rsrc_list, NULL)); */
  NODE_T *new_head = calloc_NODE_T();
  /*@ assert unchanged_ll{pre_calloc, Here}(
             to_ll{pre_calloc}(ctx->rsrc_list, NULL)); */
//@ ghost post_calloc:;  
  if (new_head == NULL){return 833;}
  /*@ assert \valid(new_head) && new_head->next == NULL; */ 
  /*@ assert ptr_sep_from_list(new_head, to_ll(ctx->rsrc_list, NULL)); */
  /*@ assert unchanged_ll{Pre, Here}(to_ll{Here}(ctx->rsrc_list, NULL));*/
//@ ghost pre_if:;
  if (ctx->rsrc_list == NULL) {
    /* The first object of the list will be added */
    ctx->rsrc_list = new_head;
    /*@ assert unchanged_ll{pre_if, Here}(to_ll(new_head, NULL));*/
    /*@ assert to_ll(new_head, NULL) == [|new_head|]; */  
    /*@ assert \separated(new_head, new_head->next);*/
    new_head->next = NULL;
    /*@ assert to_ll(new_head, NULL) == [|new_head|]; */  
  } 
  else {
    /* The new object will become the first element of the list */
    /*@ assert dptr_sep_from_list(&ctx->rsrc_list, 
                                  to_ll(ctx->rsrc_list, NULL));*/
    new_head->next = ctx->rsrc_list;
//@ ghost post_assign:;
    /*@ assert unchanged_ll{pre_if, Here}(
                            to_ll{pre_if}(ctx->rsrc_list, NULL));*/  
    /*@ assert to_ll(new_head, NULL) == 
        ([|new_head|] ^ to_ll(\at(ctx->rsrc_list, pre_if), NULL));*/
    /*@ assert dptr_sep_from_list(&ctx->rsrc_list, 
                                  to_ll(new_head, NULL));*/
    ctx->rsrc_list = new_head;
    /*@ assert unchanged_ll{post_assign, Here}(
                            to_ll{post_assign}(new_head, NULL));*/
    /*@ assert ctx->rsrc_list->next == \at(ctx->rsrc_list, Pre);*/
    /*@ assert to_ll(ctx->rsrc_list, NULL) == 
        ([|new_head|] ^ to_ll{Pre}(\at(ctx->rsrc_list, Pre), NULL));*/
    /*@ assert ctx->rsrc_list == \nth(to_ll(ctx->rsrc_list, NULL), 0);*/
  }
  //@ ghost post_add:;
  /*@ assert ctx->rsrc_list == \nth(to_ll(ctx->rsrc_list, NULL), 0);*/
  /*@ assert ctx_sep_from_list(ctx, to_ll(ctx->rsrc_list, NULL));*/
  /*@ assert ctx->rsrc_list == new_head;*/
  /*@ assert unchanged_ll{Pre, Here}(to_ll{Pre}(\at(ctx->rsrc_list, Pre), NULL));*/ 
  /*@ assert to_ll(new_head, NULL) == 
             ([|new_head|] ^ to_ll{Pre}(\at(ctx->rsrc_list, Pre), NULL));*/
  /*@ assert dptr_sep_from_list(out_node, to_ll(new_head, NULL));*/
  *out_node = new_head;
  /*@ assert unchanged_ll{post_add, Here}(to_ll{post_add}(new_head, NULL));*/
  /*@ assert ctx->rsrc_list == \nth(to_ll(ctx->rsrc_list, NULL), 0);*/
  new_head->handle = esys_handle;  
  /*@ assert \nth(to_ll(ctx->rsrc_list, NULL), 0)->handle == esys_handle;*/
  /*@ assert in_list_handle(esys_handle, to_ll(ctx->rsrc_list, NULL));*/
  /*@ assert list_sep_from_allocables(to_ll(ctx->rsrc_list, NULL)); */
  /*@ assert unchanged_ll{Pre, Here}(to_ll{Pre}(\at(ctx->rsrc_list, Pre), NULL));*/
  return 1610;
}
\end{lstlisting}
    \end{scriptsize}
    \vspace{-4mm}
    \caption{Illustrative provable example of the adjusted tpm2-tss list manipulation code, part 4/8.}
    \label{fig:self-contained-list-code-4}
\end{figure}

\begin{figure}[p]
    \begin{scriptsize}
\begin{lstlisting}[firstnumber=243,firstline=243,lastline=307,language=pretty-ACSL,basicstyle=\scriptsize\ttfamily]
/*@ 
  requires \valid(out_handle);
  assigns *out_handle;
  ensures \result \in {0, 12};
  ensures *out_handle \in {esys_handle, 0x4000000A, 0x4000000B,
                    0x40000110 + (esys_handle - 0x120U), \old(*out_handle)};
  behavior ok_handle:
    assumes esys_handle <= 31U || 0x120U <= esys_handle <= 0x12FU
         || esys_handle \in {0x10AU, 0x10BU};
    ensures \result == 0;
  behavior wrong_handle:
    assumes esys_handle > 31U 
         && (esys_handle < 0x120U || esys_handle > 0x12FU);
    assumes !(esys_handle \in {0x10AU, 0x10BU});
    ensures *out_handle == \old(*out_handle);
    ensures \result == 12;
  disjoint behaviors; complete behaviors;
*/
int iesys_handle_to_tpm_handle(uint32_t esys_handle,  uint32_t * out_handle)
{
  if (esys_handle <= 31U) {*out_handle = (uint32_t) esys_handle; return 0;}
  if (esys_handle == 0x10AU){*out_handle = 0x4000000A; return 0;}
  if (esys_handle == 0x10BU){*out_handle = 0x4000000B; return 0;}
  if (esys_handle >= 0x120U && esys_handle <= 0x12FU) 
   {*out_handle = 0x40000110 + (esys_handle - 0x120U); return 0;}
  return 12;
}

/*@ 
  requires \valid(src) && \valid(dest + (0 .. sizeof(*src)-1));
  requires \separated(dest+(0..sizeof(*src)-1),src);
  
  assigns dest[0 .. sizeof(*src)-1];
  
  ensures \valid(src);
  ensures \valid(dest + (0 .. sizeof(*src)-1));
*/
void memcpy_custom(uint8_t *dest, uint32_t *src) { 
  dest[3] = (uint8_t)(*src & 0xFF);
  dest[2] = (uint8_t)((*src >> 8) & 0xFF);
  dest[1] = (uint8_t)((*src >> 16) & 0xFF);
  dest[0] = (uint8_t)((*src >> 24) & 0xFF);
}

/*@
  requires \valid(offset) && 0 <= *offset <= UINT8_MAX - sizeof(in);
  requires buff_size > 0 && \valid(&buff[0] + (0 .. buff_size - 1));
  requires *offset <= buff_size && sizeof(in) + *offset <= buff_size;
  requires \separated(offset, buff);
  
  assigns *offset, (&buff[*offset])[0..sizeof(in) - 1];

  ensures *offset ==  \old(*offset) + sizeof(in);    
  ensures \result == 0;
*/
int uint32_Marshal(uint32_t in,uint8_t buff[],size_t buff_size,size_t *offset){ 
  size_t  local_offset = 0; 
  if (offset != NULL){local_offset = *offset;}
  in = HOST_TO_BE_32(in); 
  // memcpy(&buff[local_offset], &in, sizeof (in));
  memcpy_custom(&buff[local_offset], &in);  
  if (offset != NULL){*offset = local_offset + sizeof (in);} 
  return 0; 
}
\end{lstlisting}
    \end{scriptsize}
    \vspace{-4mm}
    \caption{Illustrative provable example of the adjusted tpm2-tss list manipulation code, part 5/8.}
    \label{fig:self-contained-list-code-5}
\end{figure}

\begin{figure}[p]
    \begin{scriptsize}
\begin{lstlisting}[firstnumber=307,firstline=307,lastline=375,language=pretty-ACSL,basicstyle=\scriptsize\ttfamily]


/*@
  requires valid_rsrc_mem_bank{Pre} && freshness(ctx, node);
  requires \valid(ctx);
  requires ctx->rsrc_list != \null ==> \valid(ctx->rsrc_list);
  requires linked_ll(ctx->rsrc_list, NULL, to_ll(ctx->rsrc_list, NULL));
  requires 0 <= \length(to_ll(ctx->rsrc_list, NULL)) < INT_MAX;
  requires \valid(node);
  requires *node != \null ==>( \valid(*node) && (*node)->next == \null);
  requires sep_from_list(ctx, node) && \separated(node, ctx);
  requires ptr_sep_from_list(*node, to_ll(ctx->rsrc_list, NULL));
  assigns _alloc_idx, _rsrc_bank[_alloc_idx], ctx->rsrc_list;
  assigns *node, (&ctx->rsrc_list->rsrc.name.name[0])[0];
  ensures valid_rsrc_mem_bank && freshness(ctx, node);
  ensures \separated(node, ctx);
  ensures \result \in {616, 833, 1611, 12};
   
  behavior handle_in_list:
    assumes in_list_handle(rsrc_handle, to_ll(ctx->rsrc_list, NULL));
    
    ensures _alloc_idx == \old(_alloc_idx);
    ensures ctx->rsrc_list == \old(ctx->rsrc_list);
    ensures in_list(*node, to_ll(ctx->rsrc_list, NULL)) && *node != NULL;
    ensures (*node)->handle == rsrc_handle && sep_from_list(ctx, node);
    ensures unchanged_ll{Pre, Post}(to_ll(ctx->rsrc_list, NULL));
    ensures \result == 616;
  behavior handle_not_converted:
    assumes !(in_list_handle(rsrc_handle, to_ll(ctx->rsrc_list, NULL)));
    assumes rsrc_handle > 31U && ! ( rsrc_handle \in {0x10AU, 0x10BU} );
    assumes rsrc_handle < 0x120U || rsrc_handle > 0x12FU;
    
    ensures unchanged_ll{Pre, Post}(to_ll(ctx->rsrc_list, NULL)); 
    ensures ptr_sep_from_list(*node, to_ll(ctx->rsrc_list, NULL));
    ensures sep_from_list(ctx, node) && *node == \old(*node);
    ensures \result == 12;
  behavior handle_not_in_list_and_node_not_allocable:
    assumes !(in_list_handle(rsrc_handle, to_ll(ctx->rsrc_list, NULL)));
    assumes rsrc_handle <= 31U || (rsrc_handle \in {0x10AU, 0x10BU})
            || (0x120U <= rsrc_handle <= 0x12FU);
    assumes _alloc_idx == _alloc_max;
    
    ensures _alloc_idx == _alloc_max;
    ensures unchanged_ll{Pre, Post}(to_ll{Pre}(ctx->rsrc_list, NULL));
    ensures *node == \old(*node) && ctx->rsrc_list == \old(ctx->rsrc_list);
    ensures ptr_sep_from_list(*node, to_ll{Pre}(ctx->rsrc_list, NULL));
    ensures sep_from_list{Pre}(ctx, node); // has to stay in behavior 
    ensures \result == 833;
  behavior handle_not_in_list_and_node_allocated:
    assumes !(in_list_handle(rsrc_handle, to_ll(ctx->rsrc_list, NULL)));
    assumes rsrc_handle <= 31U || (rsrc_handle \in {0x10AU, 0x10BU})
            || (0x120U <= rsrc_handle <= 0x12FU);
    assumes 0 <= _alloc_idx < _alloc_max;
    
    ensures in_list_handle(rsrc_handle, to_ll(ctx->rsrc_list, NULL));
    ensures (*ctx->rsrc_list).handle == rsrc_handle;
    ensures _alloc_idx == \old(_alloc_idx) + 1;
    ensures \valid(ctx->rsrc_list) && *node == ctx->rsrc_list;
    ensures ctx->rsrc_list != \old(ctx->rsrc_list);
    ensures ctx->rsrc_list->next == \old(ctx->rsrc_list);
    ensures to_ll(ctx->rsrc_list, NULL) 
        == ([|ctx->rsrc_list|] ^ to_ll{Pre}(\old(ctx->rsrc_list), NULL) );
    ensures \old(ctx->rsrc_list) != NULL ==> 
            \nth(to_ll(ctx->rsrc_list, NULL), 1) == \old(ctx->rsrc_list);
    ensures linked_ll(ctx->rsrc_list, NULL, to_ll(ctx->rsrc_list, NULL)); 
    ensures sep_from_list(ctx, node);
    ensures \result == 1611;
  disjoint behaviors; complete behaviors;
*/
\end{lstlisting}
    \end{scriptsize}
    \vspace{-4mm}
    \caption{Illustrative provable example of the adjusted tpm2-tss list manipulation code, part 6/8.}
    \label{fig:self-contained-list-code-6}
\end{figure}

\begin{figure}[p]
    \begin{scriptsize}
\begin{lstlisting}[firstnumber=376,firstline=376,lastline=416,language=pretty-ACSL,basicstyle=\scriptsize\ttfamily]
int getNode(CONTEXT *ctx, uint32_t rsrc_handle, NODE_T ** node) {
  /*@ assert linked_ll(ctx->rsrc_list, NULL, to_ll(ctx->rsrc_list, NULL));*/
  int r;
  uint32_t tpm_handle;
  { /* Block added to circumvent issues with the WP memory model */    
    NODE_T *tmp_node;
    /*@ ghost int n = 0;*/ 
    /*@ 
      loop invariant unchanged_ll{Pre, Here}(to_ll(ctx->rsrc_list, NULL));
      loop invariant linked_ll(ctx->rsrc_list, NULL, 
                     to_ll(ctx->rsrc_list, NULL));
      loop invariant linked_ll(ctx->rsrc_list, tmp_node, 
                     to_ll(ctx->rsrc_list, tmp_node));
      loop invariant ptr_sep_from_list(tmp_node,
                     to_ll(ctx->rsrc_list, tmp_node));
      loop invariant tmp_node != \null ==> 
                     in_list(tmp_node, to_ll(ctx->rsrc_list, NULL));
      loop invariant !in_list_handle(rsrc_handle, 
                     to_ll(ctx->rsrc_list, tmp_node));
      loop invariant n == \length(to_ll(ctx->rsrc_list, tmp_node));  
      for handle_in_list : loop invariant 
          in_list_handle(rsrc_handle, to_ll(ctx->rsrc_list, NULL));
      loop assigns n, tmp_node;
      loop variant \length(to_ll(tmp_node, NULL));
    */
    for (tmp_node = ctx->rsrc_list; tmp_node != NULL;
        tmp_node = tmp_node->next) {
      /*@ assert tmp_node == \nth(to_ll(ctx->rsrc_list, NULL), n);*/
      /*@ assert linked_ll(tmp_node, NULL, to_ll(tmp_node, NULL));*/
      if (tmp_node->handle == rsrc_handle){
        /*@ assert dptr_sep_from_list(node, to_ll(ctx->rsrc_list, NULL));*/
        *node = tmp_node;
        /*@ assert unchanged_ll{Pre, Here}(to_ll{Pre}(ctx->rsrc_list, NULL));*/
        /*@ assert ptr_sep_from_allocables(*node);*/
        return 616;
      }
    /*@ assert to_ll(ctx->rsrc_list, tmp_node->next) 
           == (to_ll(ctx->rsrc_list, tmp_node) ^ [|tmp_node|]);*/
    /*@ghost n++;*/   
    }  
  }
\end{lstlisting}
    \end{scriptsize}
    \vspace{-4mm}
    \caption{Illustrative provable example of the adjusted tpm2-tss list manipulation code, part 7/8.}
    \label{fig:self-contained-list-code-7}
\end{figure}

\begin{figure}[p]
    \begin{scriptsize}
\begin{lstlisting}[firstnumber=417,firstline=417,lastline=469,language=pretty-ACSL,basicstyle=\scriptsize\ttfamily]
//@ ghost post_loop:;
  /*@ assert !(in_list_handle(rsrc_handle, to_ll(ctx->rsrc_list, NULL)));*/  
  /*@ assert unchanged_ll{Pre, Here}(to_ll(ctx->rsrc_list, NULL));*/
  r = iesys_handle_to_tpm_handle(rsrc_handle, &tpm_handle);
  if (r == 12) { return r; };
  { /* Block added to circumvent issues with the WP memory model */   
    NODE_T *tmp_node_2 = NULL;
    /*@ assert dptr_sep_from_list(&tmp_node_2, 
                                  to_ll{post_loop}(ctx->rsrc_list, NULL));*/
    /*@ assert unchanged_ll{Pre, Here}(to_ll{Pre}(ctx->rsrc_list, NULL));*/
    /*@ assert \separated(node, &tmp_node_2);*/
    r = createNode(ctx, rsrc_handle, &tmp_node_2);
    /*@ assert sep_from_list(ctx, node);*/
    if (r == 833) {/*@ assert sep_from_list(ctx, node);*/ return r;};
//@ ghost post_alloc:;
    /*@ assert to_ll(ctx->rsrc_list, NULL) 
        ==([|ctx->rsrc_list|] ^ to_ll{Pre}(\at(ctx->rsrc_list, Pre), NULL));*/
    /*@ assert ctx_sep_from_list(ctx, to_ll(ctx->rsrc_list, NULL));*/
    tmp_node_2->rsrc.handle = tpm_handle;
    tmp_node_2->rsrc.rsrcType = 0;  
    size_t offset = 0;
    /*@ assert ptr_sep_from_list(tmp_node_2,
               to_ll(ctx->rsrc_list->next, NULL));*/
    r = uint32_Marshal(tpm_handle, &tmp_node_2->rsrc.name.name[0],
                        sizeof(tmp_node_2->rsrc.name.name),&offset);     
    /*@ assert in_list_handle(rsrc_handle, to_ll(ctx->rsrc_list, NULL));*/
    if (r != 0) { return r;};
    tmp_node_2->rsrc.name.size = offset;
    /*@ assert unchanged_ll{post_alloc, Here}(to_ll(ctx->rsrc_list, NULL));*/
    /*@ assert dptr_sep_from_list(node, to_ll(ctx->rsrc_list, NULL));*/
    /*@ assert dptr_sep_from_list(node, 
               to_ll{Pre}(\at(ctx->rsrc_list, Pre), NULL));*/
    *node = tmp_node_2;
    /*@ assert unchanged_ll{Pre, Here}(
               to_ll{Pre}(\at(ctx->rsrc_list, Pre), NULL));*/
  }
  /*@ assert unchanged_ll{Pre, Here}(
             to_ll{Pre}(\at(ctx->rsrc_list, Pre), NULL));*/
  /*@ assert in_list_handle(rsrc_handle, to_ll(ctx->rsrc_list, NULL));*/
  /*@ assert ctx->rsrc_list->next == \at(ctx->rsrc_list, Pre);*/
  /*@ assert \at(ctx->rsrc_list, Pre) != \null ==> 
      ctx->rsrc_list->next == \nth(to_ll(ctx->rsrc_list, NULL), 1);*/
  /*@ assert ctx->rsrc_list->handle == rsrc_handle;*/
  /*@ assert freshness(ctx, node);*/
  /*@ assert sep_from_list(ctx, node);*/
  return 1611;
}

/* Command to run the proof with Frama-C:
frama-c-gui -c11 example.c -wp -wp-rte -wp-prover altergo,cvc4,cvc4-ce,script 
-wp-timeout 50 -wp-smoke-tests -wp-prop="-@lemma"
*/
\end{lstlisting}
    \end{scriptsize}
    \vspace{-4mm}
    \caption{Illustrative provable example of the adjusted tpm2-tss list manipulation code, part 8/8.}
    \label{fig:self-contained-list-code-8}
\end{figure}

\end{document}